\documentclass[10pt,prl,aps,superscriptaddress,amsmath,amssymb,twocolumn,floatfix,longbibliography,footinbib]{revtex4-1}
\usepackage{mathtools}
\usepackage{mathrsfs, amsmath, wasysym}
\usepackage{bbold}
\usepackage{dsfont}
\usepackage{braket,mleftright}
\usepackage[mathscr]{eucal}
\usepackage{xfrac}
\usepackage{graphicx}
\usepackage{dcolumn}
\usepackage{bm}
\usepackage{color}
\usepackage{tabularx}
\usepackage{natbib}
\usepackage[bookmarksopen, colorlinks,linkcolor=blue,citecolor=blue,urlcolor=blue]{hyperref}
\usepackage[normalem]{ulem}
\usepackage[all]{hypcap}
\usepackage[dvipsnames,usenames,table]{xcolor}

\newcommand{\bg}{\begin{gather}}
\newcommand{\eg}{\end{gather}}

\newcommand{\be}{\begin{equation} }
\newcommand{\ee}{\end{equation}}
\newcommand{\beq}{\begin{eqnarray}}
\newcommand{\eeq}{\end{eqnarray}}

\newcommand{\SO}{\text{SO}}

\begin{document}

\title{Mirror Chern Bands and Weyl Nodal Loops in Altermagnets}

\author{Daniil S. Antonenko}
\affiliation{Department of Physics, Yale University, New Haven, Connecticut 06520, USA}

\author{Rafael M. Fernandes}
\affiliation{School of Physics and Astronomy, University of Minnesota, Minneapolis, Minnesota 55455, USA}

\author{J{\"o}rn W. F. Venderbos}
\affiliation{Department of Physics, Drexel University, Philadelphia, Pennsylvania 19104, USA}
\affiliation{Department of Materials Science \& Engineering, Drexel University, Philadelphia, Pennsylvania 19104, USA}

\date{\today}

\begin{abstract}
The electronic spectra of altermagnets are a fertile ground for nontrivial topology due to the unique interplay between time-reversal and crystalline symmetries. This is reflected in the unconventional Zeeman splitting between bands of opposite spins, which emerges in the absence of spin-orbit coupling (SOC) and displays nodes along high-symmetry directions. Here, we argue that even for a small SOC, the direction of the magnetic moments in the altermagnetic state has a profound impact on the electronic spectrum, enabling novel topological phenomena to appear. By investigating microscopic models for two-dimensional (2D) and three-dimensional (3D) altermagnets, motivated in part by rutile materials, we demonstrate the emergence of hitherto unexplored Dirac crossings between bands of the same spin but opposite sublattices. The direction of the moments determines the fate of these crossings when the SOC is turned on. We focus on the case of out-of-plane moments, which forbid an anomalous Hall effect and thus ensure that no weak magnetization is triggered in the altermagnetic state. In the 2D model, the SOC gaps out the Dirac crossings, resulting in mirror Chern bands that enable the Quantum Spin Hall Effect and undergo a topological transition to trivial bands upon increasing the magnitude of the magnetic moment. On the other hand, in the 3D model the crossings persist even in the presence of SOC, forming Weyl nodal loops protected by mirror symmetry. Finally, we discuss possible ways to control these effects in altermagnetic material candidates.  
\end{abstract}

\maketitle

{\it Introduction---}A defining property of a collinear magnetic configuration that does not display a macroscopic magnetization is that it must remain invariant when time-reversal (TR) symmetry is combined with an additional crystalline symmetry \cite{Smejkal2020,Smejkal-AltermagnetismPerspective,Kusunose2019,Kunes2019,Zunger2020,Kusunose2020,Turek2022,Spaldin2022}. When this additional symmetry involves a rotation (proper or improper), as proposed for a wide range of materials \cite{Smejkal-Altermagnetism,Mazin2021,Smejkal2022chiral,Facio2023,Autieri2023,Mazin2023,Lovesey2023,Autieri2023_perovskites,Mazin2023_FeSe,Aoyama2023,Zhu2023,Gao2023,Grzybowski2023,Smolyanyuk2023,Kawamura2023,Sodequist2024}, the system is classified as an altermagnet (AM) rather than an antiferromagnet, in which case the additional symmetry is either translation or inversion \cite{Smejkal-AltermagnetismPerspective,Smejkal-AltermagnetismReview}. In contrast to collinear antiferromagnets, the band structure of an AM displays a nonzero spin splitting (also called Zeeman splitting), whose magnitude can be as large as 1 eV \cite{Smejkal-AltermagnetismPerspective,Facio2023}. Unlike ferromagnets, the Zeeman splitting is not uniform, but instead has a nodal momentum-space structure (e.g. $d$-wave) \cite{Smejkal-Altermagnetism} similar to metals undergoing certain Pomeranchuk instabilities \cite{Fradkin2007,Pomeranchuk1958}, which can be exploited for spintronic applications \cite{Smejkal-AltermagnetismPerspective,Bai2023,Zhang2023,Chi2023,Sun2023_spin,Ang2023,Hodt2023} or superconducting heterostructures \cite{Ouassou2023,Neupert2023,Beenakker2023,Sun2023,Papaj2023,Zhu2023,Cano2023,chakraborty2023,Li2023}. Importantly, evidence for altermagnetism has been reported in various materials \cite{Feng-RuO2AltermagnetismExperiment,Betancourt2023,Zhou2023,Fedchenko2023,Lee2023,Osumi2023,Kluczyk2023,Reimers2023, lin2024observation}.   

While a negligible spin-orbit coupling (SOC) is often assumed \cite{Smejkal-Altermagnetism}, the latter is not incompatible with altermagnetism \cite{Fernandes-Altermagnetism,Steward2023,Agterberg2023,Wei2023,Galindez2023,Hariki2023,Fakhredin2023,JCano-QuantumGeometryAltermagnetism}. On the contrary, important phenomena inside the AM phase are enabled even by a small SOC, such as the splitting of the Zeeman nodal planes into nodal lines \cite{Fernandes-Altermagnetism} and
the emergence of the anomalous Hall effect (AHE) \cite{Smejkal2020,Feng-RuO2AltermagnetismExperiment,Betancourt2023,Kluczyk2023} accompanied by weak ferromagnetism \cite{MacDonald2020,Tsymbal2020}. The latter occurs only for particular orientations of the AM magnetic moments, showcasing the influence of the moments direction on the AM behavior.

\begin{figure}
	\includegraphics[width=0.98\columnwidth]{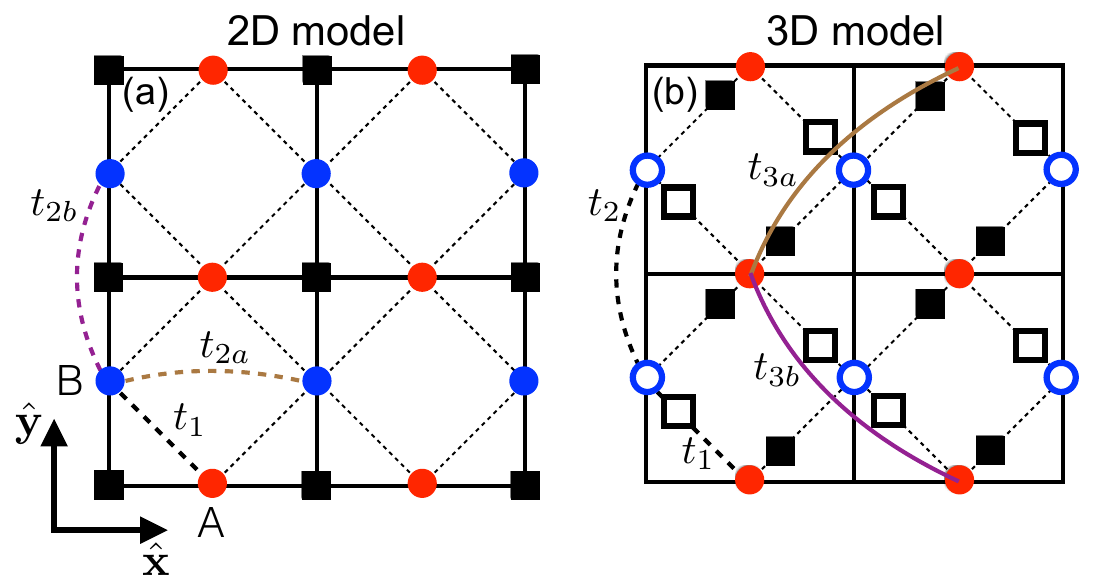}
	\caption{\label{fig:lattice} (a) Lieb-lattice model for altermagnetism in 2D. Red and blue dots (i.e., $A$ and $B$ sublattice) denote antialigned moments; the crystalline environment provided by the nonmagnetic sites (black squares) generates anisotropic second-nearest-neighbor hopping $t_{2a,b}$. (b) Rutile-lattice model for altermagnetism in 3D (top view). Third-nearest neighbor hoppings $t_{3a,b}$ are anisotropic due to the presence of the nonmagnetic sites. Open blue dots and black squares denote sites in the $z=1/2$ plane. }
\end{figure}

In this Letter, we demonstrate the existence of additional band crossings in the phase diagram of model altermagnets, which are distinct from the nodal Zeeman splitting. By investigating paradigmatic microscopic AM models in two and three dimensions (2D and 3D), see also \cite{Sudbo-minimal, McClarty_LongSpinGroups,McClarty-AltermagnetismGL}, corresponding, respectively, to altermagnets with $d_{x^2-y^2}$ and $d_{xy}$ spin splitting, we show that, in the absence of SOC, bands of the same spin but opposite sublattice character cross and realize Dirac points (2D) and Weyl nodal lines (3D)~\cite{Fang:2016p117106,Weng:2016p303001,Armitage:2018p015001,Bernevig:2018p041001,Hirayama:2018p041002,Gao:2019p153} protected by mirror symmetries. The fate of these additional band crossings in the presence of SOC depends crucially on the magnetic moments direction. Focusing on the less explored case of moments along the $z$ direction, which forbids AHE, the Dirac points of the 2D altermagnet are gapped by SOC, giving rise to a mirror Chern insulator. Therefore, rotating the magnetic moments of a 2D AM from an in-plane to the out-of-plane direction  ``converts'' the AHE into the Quantum Spin Hall Effect (QSHE).  In contrast, the Weyl line nodes in 3D are robust against SOC. Hence, our work uncovers previously unexplored topological characteristics of altermagnets and elucidates the crucial role of the magnetic moments' direction.

{\it Model altermagnet in 2D---}A simple realization of an altermagnet in 2D is a variant of the Lieb lattice shown in Fig. \ref{fig:lattice}, where the antiparallel magnetic moments (blue and red dots) form two distinct sublattices related by fourfold rotation. Importantly, the $A$ and $B$ sublattices experience distinct local environments due to the nonmagnetic site (black squares). 
Here we consider a Kondo-type model with frozen localized spins and itinerant electrons hopping only among the two magnetic sublattices. It corresponds to a projection of the three-band model studied in Ref. \cite{Sudbo-minimal} in the absence of SOC~\cite{SM}.

The itinerant electrons have a Kondo coupling $J$ to the collinear local moments \cite{Smejkal2020,McClarty_LongSpinGroups,Sudbo-minimal,Okamoto2023}, which have staggered magnetization $\bm{N} = \bm{M}_A - \bm{M}_B$. We introduce the electron operators $c_{\bm{k},s\mu}$ with momentum $\bm{k}$, spin projection $s=\uparrow, \, \downarrow$ and sublattice $\mu=A,B$, defining $\bm{c}_{\bm{k}} = \left(c_{\bm{k},\uparrow A}, \, c_{\bm{k},\uparrow B}, \, c_{\bm{k},\downarrow A}, \, c_{\bm{k},\downarrow B} \right)^T$. 
The Hamiltonian $\mathcal{H}_0(\bm{k})$ without SOC contains both nearest-neighbor (NN) hopping $t_1$ and anisotropic next-nearest-neighbor (NNN) hopping $t_{2a}$ and $t_{2b}$, due to the different crystallographic environments of the $A$ and $B$ sublattices (see also \cite{Sudbo-minimal}):
\begin{multline} \label{hamiltonian}
	\mathcal{H}_0(\bm{k}) =
	 - 4t_{1}\cos\frac{k_{x}}{2}\cos\frac{k_{y}}{2}\tau_{x}-2t_{2}(\cos k_{x}+\cos k_{y})\tau_0  \\
	 - 2t_{d}(\cos k_{x}-\cos k_{y})\tau_{z}
	 + J \tau_{z} \bm{N} \cdot \bm{\sigma}.
\end{multline}
Here, $\bm{\sigma}$ and $\bm{\tau}$ are Pauli matrices in the spin and sublattice spaces, respectively, and $t_{2,d} = (t_{2a} \pm t_{2b})/2$. 

\begin{figure}
	\includegraphics[width=0.99\columnwidth]{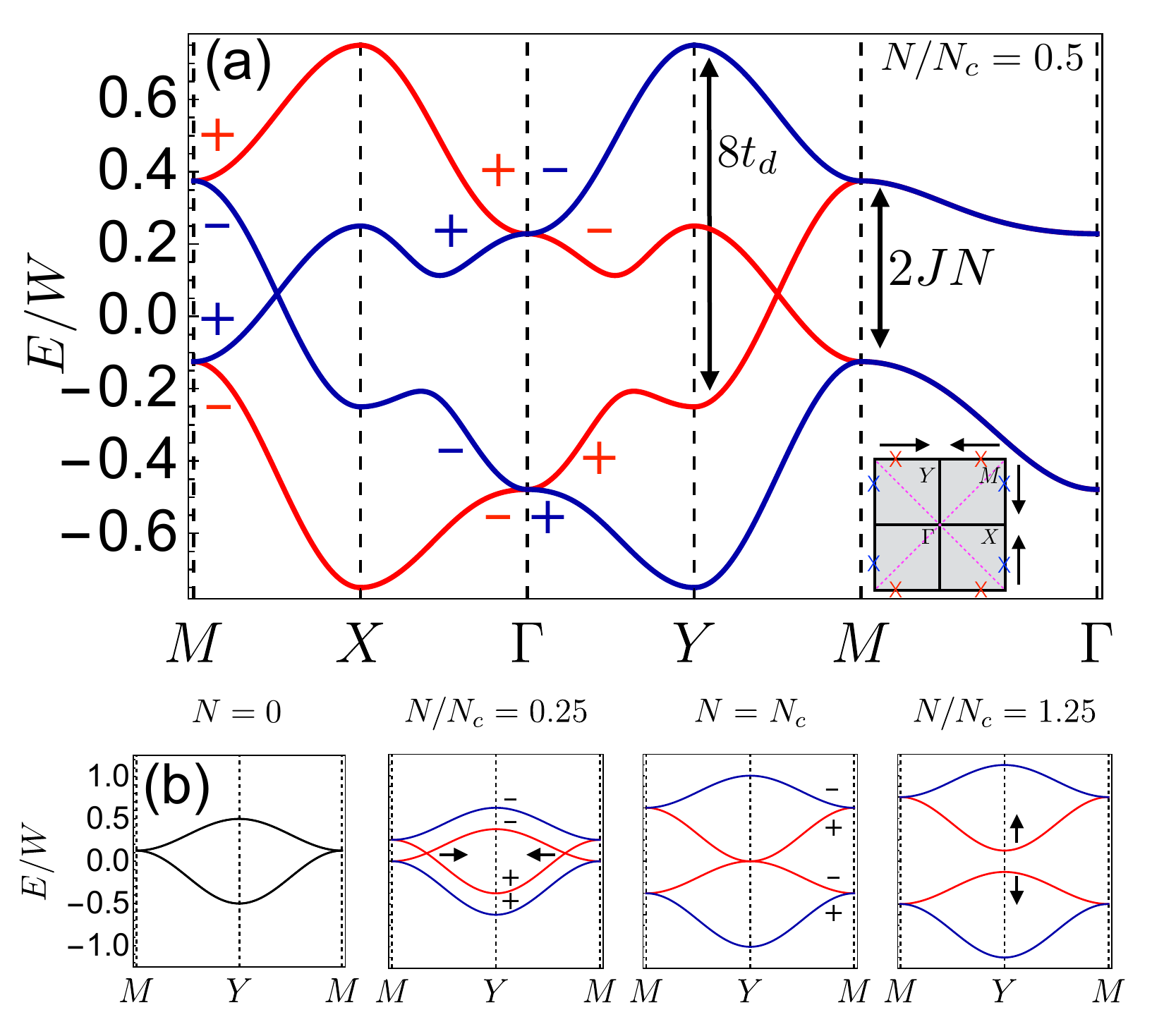}
	\caption{\label{fig:bands} (a) Electronic dispersion of the 2D altermagnetic model with spin-up and spin-down bands shown in red and blue, respectively. Dirac crossings between bands of the same spin occur on the BZ edges (see inset). The values $t_2/t_1 = 0.5$, $t_d / t_1 = 2$ were used, and $W \equiv 8 t_d$.
	(b) Band dispersion along the $MY$ line as a function of the N\'eel vector amplitude $N$. From left to right: the quadratic band crossing at $M$ splits into linear Dirac crossings between bands of opposite mirror $\mathcal{M}_y$ eigenvalues $\pm$, which merge at $Y$ and are removed at $N=N_c \equiv |4 t_d / J|$. }
\end{figure}

Without SOC, the electron spin quantization axis is along the arbitrary direction $\bm{N}$, resulting in a block diagonal $\mathcal{H}_0$ with $J \tau_{z} \bm{N} \cdot \bm{\sigma} \rightarrow \pm JN \tau_z$. 
The typical band dispersion is shown in Fig.~\ref{fig:bands}(a). If $t_d = 0$, the model reduces to a simple square-lattice antiferromagnet, which has degenerate spin bands. A nonzero $t_d$ lifts the spin degeneracy everywhere except along the diagonal $k_x = \pm k_y$ lines, resulting in a Zeeman splitting with $d_{x^2-y^2}$-wave symmetry. Note that nodes in the Zeeman splitting can also emerge in an antiferromagnet subjected to external magnetic fields \cite{Ramazashvili2008}.

Besides nodal lines, the spectrum in Fig.~\ref{fig:bands}(a) also exhibits band crossings on the Brillouin zone edges ($MX$ and $MY$ lines). These are crossings between bands of the same spin but different sublattice, not pinned to a particular point in the high-symmetry line. Indeed, setting $\bar{\bm{k}}=(k,\pi)$ along the $MY$ line leads to a diagonal $\mathcal{H}_0(\bar{\bm{k}}) = 4t_{2}\sin^2(k/2)\tau_0-[4t_{d}\cos^2 (k/2) \mp JN]\tau_{z}$. Thus, degeneracies occur when $\cos^2 (k/2) =\pm JN/4t_d$, which implies a crossing of spin-$\uparrow$ or spin-$\downarrow$ bands, provided $|JN/4t_d | \leq 1$. Along the $MX$ line, the spin of the crossing bands is opposite to that along $MY$, reflecting the $d$-wave symmetry of the altermagnet. The evolution of these band crossings is shown in Fig.~\ref{fig:bands}(b). They are created at the $M$ point for infinitesimal $N$, move towards $X$ ($Y$) as $N$ increases, and are eventually removed at $X$ ($Y$) when $N = N_c \equiv |4t_d/J|$. 

In general, band crossings are protected when the bands involved have different eigenvalues under a crystal symmetry. Without SOC, the mirror reflections $\mathcal M_x:x\rightarrow -x$ and $\mathcal M_y:y\rightarrow -y$ are good symmetries, since they do not exchange sublattices and leave $MX$ and $MY$ invariant, respectively. 
Indeed, $\mathcal M_y$ maps a $B$ ($A$) site onto another $B$ ($A$) site related by an odd (even) number of elementary translations $T({\hat y})$. Since $k_y=\pi$ on the $MY$ line, the mirror eigenvalues are equal to $-1$ ($+1$) for the $B$ ($A$) sublattice, leading to a protected crossing between bands of same spin but different sublattices.

This protection is also manifested when expanding $\mathcal{H}_0(\bm{k})$ in small $\bm{q}$ around $\bm{Q}_M=(\pi,\pi)$:
\begin{equation} \label{M_expansion}
	\mathcal{H}^{\uparrow,\downarrow}_0(\bm{q}+\bm{Q}_M) =  \epsilon_q \tau_0 - \left[ t_2 (q_x^2 - q_y^2) \mp J N \right] \tau_z  
	 - t_1 q_x q_y \tau_x  ,
\end{equation}  
where $\epsilon_q =t_2 (4-q^2)$. Thus, $\mathcal{H}^{\uparrow,\downarrow}_0$ takes the form of a quadratic band crossing (QBC) Hamiltonian \cite{FradkinKivelson} with symmetry-breaking perturbation $\pm J N \tau_z$, which, in Ref.~\onlinecite{FradkinKivelson}, was referred to as ``nematic-spin-nematic''. 
When $|JN|>0$, the QBC splits into two sets (one for each spin) of two Dirac points located on the $q_x=0$ and $q_y=0$ lines. 
Therefore, altermagnetism on the two-sublattice square lattice generically leads to Dirac crossings on mirror-symmetric lines in the absence of SOC.

{\it Model 2D altermagnet and SOC---}Including SOC in the tight-binding model of Eq.~\eqref{hamiltonian} gives rise to the additional term 
\be \label{SO_hamiltonian}
\mathcal{H}_\SO (\bm{k}) = \lambda \sin \frac{k_x}{2} \sin \frac{k_y}{2}\tau_y \sigma_z, 
\ee
which is of Kane-Mele type \cite{KaneMele-QSHE} and is allowed by all lattice symmetries \cite{Franz2010}. Because the N\'eel vector $\bm{N}$ can no longer rotate independently of the lattice, one must distinguish in-plane vs. plane-normal directions of $\bm{N}$. 

\begin{figure}
	\includegraphics[width=0.95\columnwidth]{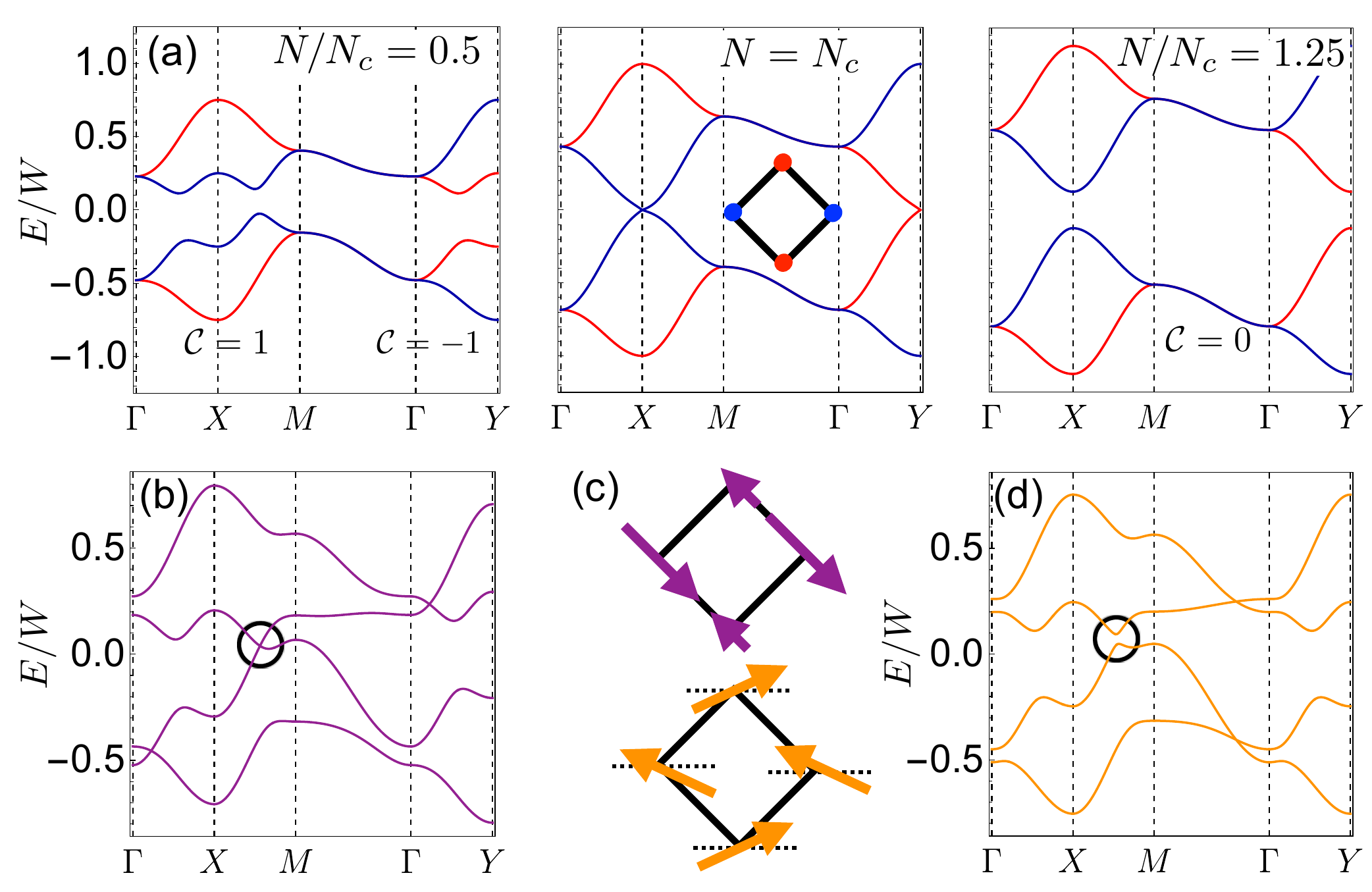}
    \caption{\label{fig:soc-bands} 
 Impact of SOC on the electronic dispersion of the 2D altermagnet shown in Fig.~\ref{fig:bands}.
 (a) For an out-of-the-plane N\'eel vector, a topological transition to a trivial state takes place upon increasing $N$ at $N = N_c$ (left to right panels). Here, $\lambda / t_1 = 2$. Panels (b) and (d) show the cases of an in-plane N\'eel vector of magnitude $N = 0.5 N_c$ oriented along the principal axis (b) and the diagonal (d). The ferrimagnetic and canted magnetic configurations resulting from the mixture with the induced ferromagnetic moment are shown in the upper and lower panels of (c), respectively. The values $\lambda / t_1 = 3$ and $\lambda_{\text{FM}} N / t_1 = 0.7$ were used (see Supplemental Material for the definition of $\lambda_{\text{FM}}$).
  }
\end{figure}

Consider first $\bm{N} = N \hat z$, for which secondary ferromagnetism (and thus AHE) is not allowed~\cite{SM}. In this case the system maintains a horizontal mirror symmetry $\mathcal M_z$ with Bloch representation $U_{\mathcal M_z} = -i \sigma_z$, which renders the Hamiltonian block-diagonal in spin-space. To understand the effect of \eqref{SO_hamiltonian}, it is instructive to again expand it around $M$, $\mathcal{H}^{\uparrow,\downarrow}_\SO (\bm{q}+\bm{Q}_M) = \pm \lambda \tau_y$. Clearly, this term couples the two sublattices within the same spin sector, thus gapping the Dirac points of $\mathcal{H}_0$ in Eq.~\eqref{M_expansion}. Indeed, the mirror reflections $\mathcal M_{x(y)}$ that protected the crossings are no longer symmetries in the presence of SOC, as they do not leave the spins invariant. The resulting band structure is shown in Fig.~\ref{fig:soc-bands}(a) for parameters that yield an insulator at half filling. 

Gapping the Dirac points associated with a QBC is known to produce Chern bands~\cite{FradkinKivelson}. Here, they are produced in each spin sector but with opposite Chern numbers $C=\pm 1$. Since the spin sectors are identified with the $\mathcal M_z$ eigenvalues $\pm i$, the half-filled insulator realizes a mirror Chern insulator protected by mirror Chern number $C_{\mathcal M} = (C_{+i} - C_{-i})/2$~\cite{Hsieh:2012p982}. Sample boundaries must then host a pair of chiral edge states with opposite chirality for each mirror/spin sector, resulting in a Quantum Spin Hall Effect~\cite{KaneMele-QSHE}, provided $|J N / 4 t_d| < 1$. As shown in Fig.~\ref{fig:soc-bands}(a), as $N$ increases towards $N_c$, a gap closing transition occurs at $X$ and $Y$ described by a Dirac fermion mass inversion \cite{SM}, which renders the gap trivial for $N > N_c$. See the Supplemental Material \cite{SM} for an explicit demonstration of the Berry curvature and Quantum Spin Hall Effect, utilizing Refs.~\cite{Fukui-Berry, Vanderbilt-BerryBook, Kwant-package}.  

Next, consider the case of in-plane N\'eel vector $(N_x, N_y)=N(\cos \alpha, \, \sin \alpha)$. Below $T_N$, its orientation $\alpha$ is determined by the Landau free energy $F_N = a_N N^2 + u_N N^4 + \gamma_N N^4 \cos 4\alpha $, which is minimized by $\alpha = 0, \pi/2$ when $\gamma_N<0$ and $ \alpha = \pm \pi/4$ when $\gamma_N > 0$. Importantly, for either case, the horizontal mirror $\mathcal M_z$ is broken and the Hamiltonian does not have a manifest block diagonal structure. Furthermore, an overall ferromagnetic moment (i.e., a uniform magnetization) $\bm{M} = \bm{M}_A+\bm{M}_B = M(\cos \eta, \, \sin \eta)$ is induced in the AM state \cite{Fernandes-Altermagnetism} due to the symmetry-allowed coupling $F_{N,M} = \kappa N M \cos (\alpha + \eta)$. Consequently, a new Hamiltonian term $H_{\text{FM}} = \lambda_{\text{FM}} (N_x \sigma_x - N_y \sigma_y)$ emerges, generating itinerant ferromagnetism.

Thus, when the N\'eel vector is oriented along the principal axes ($\alpha = 0, \pi/2$), the induced $\bm{M}$ is parallel to $\bm{N}$. This results in a ferrimagnetic state [upper panel of Fig.~\ref{fig:soc-bands}(c)] that retains, for $\alpha=0$ ($\alpha=\pi/2$), the mirror symmetry $\mathcal M_x$ ($\mathcal M_y$) and the rotation symmetry $\mathcal C_{2x}$ ($\mathcal C_{2y}$), which leave the $MX$ and $MY$ ($MY$ and $MX$) lines invariant, respectively. Thus, as shown in Fig.~\ref{fig:soc-bands}(b), the Dirac points described by Eq.~\eqref{M_expansion} remain protected, and can only be removed for a large enough SOC coupling $\lambda$ in Eq.~\eqref{SO_hamiltonian} (see Supplemental Material). In contrast, these mirror and rotation symmetries are broken when the N\'eel vector is oriented along the diagonals ($\alpha = \pm \pi/4$), since the induced $\bm{M}$ is orthogonal to $\bm{N}$, resulting in a noncollinear canted state [lower panel of Fig.~\ref{fig:soc-bands}(c)]. This allows the crossing bands to mix, gapping the Dirac points [Fig.~\ref{fig:soc-bands}(d)].

\begin{figure}
	\hspace{-0.6cm}
		\includegraphics[width=0.98\columnwidth]{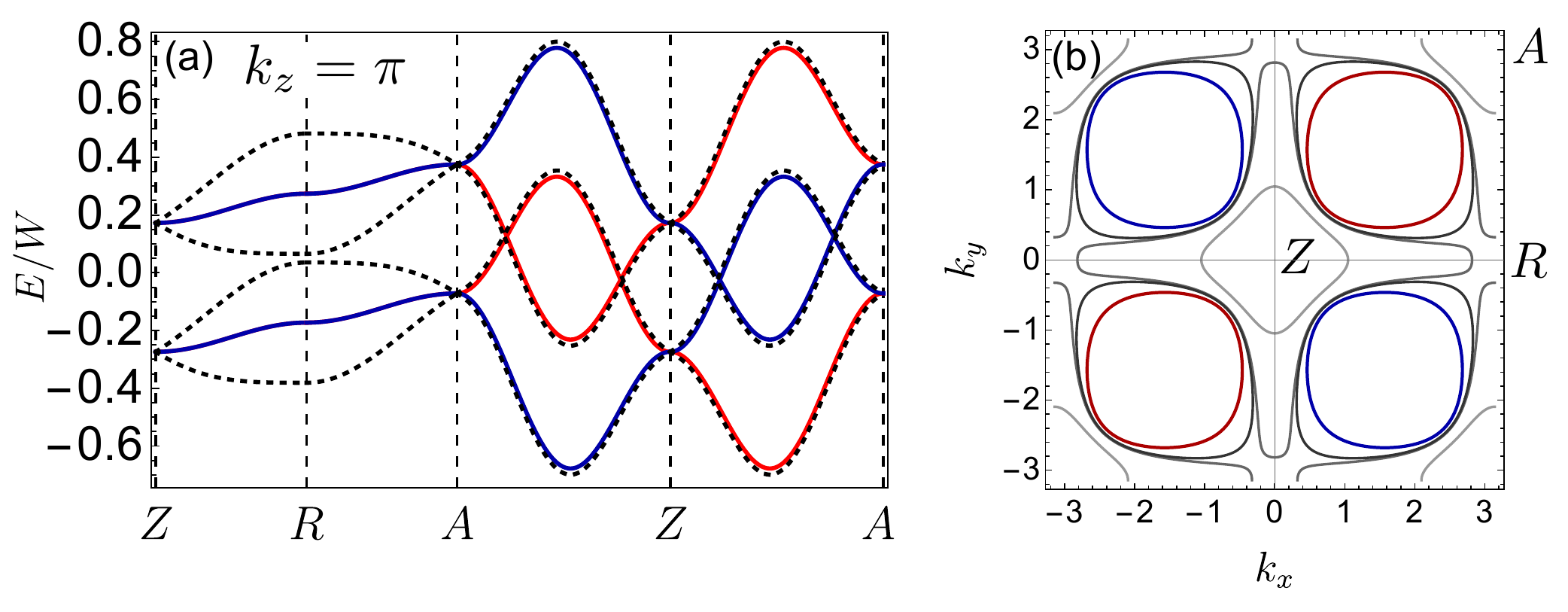}
	\caption{\label{fig:3D} 
Band structure of the 3D rutile model on the $k_z = \pi$ plane with $\bm{N} \parallel \bm{\hat{z}}$ for $t_2 / t_1 = 0.85$, $t_d / t_1 = 4.2$, $JN / t_1 = 7.5$ plotted in the units of $W=8t_d$. (a) Energy bands without SOC [red(blue) for spin-up(-down)] and with SOC (dashed black), with $\lambda_2 / t_1 = 7$. (b) Evolution of the Weyl nodal lines on the $k_z = \pi$ plane from $\lambda_2=0$ (no SOC, red and blue) to $\lambda_2 = 15 t_1$ (thinnest gray lines). 
 }
\end{figure}

{\it Model altermagnet in 3D---}We now turn to altermagnets in 3D, focusing on a microscopic model relevant for AM rutile materials, such as RuO$_2$ and MnF$_2$ \cite{Smejkal-Altermagnetism,Spaldin2022}. Described by space group $P4_2/mnm$ (\#136), they form a two-sublattice structure in which the magnetic atoms (Ru or Mn) sit at Wyckoff positions $(0,0,0)$ and $(\frac12,\frac12,\frac12)$ [see Fig.~\ref{fig:lattice}(b)]. As in the 2D model, the arrangement of the nonmagnetic sites (black squares) leads to different crystallographic environments on the magnetic $A$ and $B$ sites, whose moments are antialigned. The sublattices are exchanged by nonsymmorphic operations, namely, fourfold and twofold rotations followed by a half-translation along the diagonal, $\{ \mathcal C_{4z}|\frac12\frac12\frac12 \}$ and $\{ \mathcal C_{2x}|\frac12\frac12\frac12 \}$. Following Refs.~\cite{McClarty_LongSpinGroups,McClarty-AltermagnetismGL,Okamoto2023}, we consider
\begin{multline} \label{ham-3D}
	\mathcal{H}_0(\bm{k}) =
	 - 8t_{1}c_{x/2}c_{y/2}c_{z/2} \tau_{x}-2t_{2}(c_x + c_y)\tau_0   \\
	-2t'_{2}c_z\tau_0 - 4t_{d}s_x s_y\tau_{z} + J \tau_{z} \bm{N} \cdot \bm{\sigma},
\end{multline}
where we defined $c_{i}\equiv \cos k_i$, $s_{i}\equiv \sin k_i$,  $c_{i/2} \equiv \cos{k_i /2}$, $s_{i/2} \equiv \sin{k_i/2}$, and $\tau_z = \pm 1$ corresponds to the $A$ and $B$ sublattice. Here, $t_1$, $t_2$ ($t'_2$), and $t_d = t_{3a} - t_{3b}$ are first-, second-, and anisotropic third-nearest neighbor hopping, respectively (we ignore physically unimportant symmetric third-nearest hopping). Similar to the 2D model, $t_d$ reflects the distinct crystallographic environment of the $A$ and $B$ sites, producing a nodal spin splitting with $k_xk_y$ symmetry (i.e. nodal planes) ~\cite{McClarty_LongSpinGroups,McClarty-AltermagnetismGL}. With SOC, two additional terms emerge \cite{SM}
\begin{multline} \label{SO_ham-3D}
\mathcal{H}_\SO (\bm{k})  =
	\lambda_1 c_{x/2}c_{y/2}c_{z/2} (c_x-c_y)\tau_y \sigma_z  \\
+	\lambda_2 s_{z/2} (s_{x/2}c_{y/2}\tau_y \sigma_x - s_{y/2}c_{x/2}\tau_y \sigma_y) ,
\end{multline}
enabling us to establish generic and robust electronic properties of rutile altermagnets in the presence of SOC.
While the case of an in-plane N\'eel vector ($\bm{N} \perp \bm{\hat{z}}$) has been widely studied \cite{Smejkal-AltermagnetismPerspective}, as it displays an AHE, here we focus on $\bm{N} = N \bm{\hat z}$, for which both a Hall vector and ferromagnetism are forbidden. 

Our analysis reveals three general properties: (i) the Zeeman-splitting nodal planes are lifted, leaving behind only nodal lines along high-symmetry directions, in agreement with Ref.~\cite{Fernandes-Altermagnetism}. (ii) The vertical Zeeman Weyl nodal lines that coincide with the BZ boundaries can be understood as the (alter)magnetically split Dirac nodal lines of the nonmagnetic state uncovered in Ref.~\cite{Sun:2017p235104}. (iii) The $k_z=\pi$ plane generically hosts horizontal Weyl nodal lines, corresponding to a crossings of bands of opposite mirror $\mathcal M_{z}$ eigenvalues. Since these line nodes are fundamentally different from the widely studied $d$-wave Zeeman splitting, we hereafter focus on them only. A detailed discussion of properties (i) and (ii) is presented in the Supplemental Material~\cite{SM}.

Without SOC, all bands on the $k_z = \pi$ plane can be labeled by their spin projection [red(blue) for up(down)], see Fig.~\ref{fig:3D}(a). Besides the usual crossing between bands of opposite spin along the main axes ($ZR$ and $RA$ lines), which correspond to the $d$-wave Zeeman nodal lines, bands of same spin but opposite sublattice cross along the diagonals ($ZA$ direction). Like the 2D case, these crossings are protected as long as $|JN /4 t_d|<1$, since the bands have different mirror $\mathcal M_{z}$ eigenvalues. When the SOC term $\lambda_2$ (the only nonzero SOC term on the $k_z=\pi$ plane) is included (dashed lines), although spin is no longer a good quantum number, these crossings remain protected as $\mathcal M_{z}$ acts trivially on $\bm{N} = N \bm{\hat z}$ and thus remains a symmetry. In contrast, the crossings between bands of opposite spins along $ZR$ and $RA$ are completely gapped, in agreement with Ref. \cite{JCano-QuantumGeometryAltermagnetism}. As shown in Fig.~\ref{fig:3D}(b), the crossings between same-spin bands actually form four Weyl nodal loops along the $k_z = \pi$ plane enclosing the midpoints of the $ZA$ diagonals (red and blue loops). Interestingly, upon increasing $\lambda_2$, these Weyl line nodes eventually undergo a Lifshitz transition, forming two loops enclosing the $Z$ and $R$ points. The behavior in the limit of strong SOC can be understood from the fact that the $Z$ point realizes a symmetry-enforced Dirac point in the nonmagnetic state, which is energetically split into Weyl points by a nonzero $N$.  

{\it Conclusion---}In summary, our work reveals the existence of topological nodal crossings between bands of same spin but opposite sublattice in the phase diagram of minimal models of 2D and 3D altermagnets. These crossings do not require fine-tuning and appear generically for nonzero $JN$, a result most clearly demonstrated by the continuum $k\cdot p$ theories developed around the $M$ point (in the 2D model) and the $A$ point (in the 3D model, see Supplemental Material~\cite{SM}). These degeneracies are fundamentally different from those associated with the widely studied Zeeman nodal lines characteristic of altermagnets. Importantly, the orientation of the magnetic moments determines the fate of these crossings in the presence of SOC. In 2D, out-of-plane moments gap the Dirac crossings along the BZ edges and give rise to Chern bands, thus opening an interesting route to realize mirror Chern insulating altermagnets. In contrast, in 3D, the Weyl nodal lines remain robust against the presence of SOC. 

Since the 3D model studied here is based on the rutile structure, the predicted mirror-protected Weyl nodal lines along the $k_z = \pi$ plane should be present in the altermagnet candidates RuO$_2$ and MnF$_2$. Spin-resolved spectroscopy, which can be performed via ARPES or STM measurements, could detect these features by directly probing the electronic dispersion. In the 2D case, the manifestations of the mirror Chern bands include the occurrence of edge states and of the Quantum Spin Hall Effect. Note that two 2D altermagnet candidates are described by a Lieb lattice: the Janusized monolayer Fe(Se,Te) \cite{Mazin2023_FeSe} and monolayer Cr$_2$Te$_2$O \cite{Cui2023,Bai2024}. Interestingly, artificial Lieb lattices were implemented in covalent organic framework (COF) systems \cite{OrganicLieb-Liu, OrganicLieb-Huang}, some of which display magnetism.

Finally, in either the 3D or the 2D case, the band crossings from which the topological phenomena emerge can be removed by increasing the ratio $J N/t_d$. This offers two different tuning knobs to drive the altermagnetic band structure across a topological transition: temperature, which indirectly controls the size of the staggered magnetization $N$, and strain or pressure, which should change $t_d$ by altering the ligand fields responsible for the two-sublattice structure \cite{Wang_Gautreau}. Additionally, changing the size of the nonmagnetic atoms should also affect $t_d$, as the latter arises from virtual hopping between the magnetic atoms and the nonmagnetic atoms (see section II.D in the Supplemental Material \cite{SM}).

\begin{acknowledgments}
{\it Acknowledgments---}We thank J. Cano and S. Ghorashi for useful discussions, and L. \v{S}mejkal for bringing to our attention altermagnet candidates with a Lieb lattice structure. D.S.A. was supported by NSF Grant No. DMR-2002275 and Yale Mossman Fellowship. J.W.F.V. was supported by the National Science Foundation under Award No. DMR-2144352. R.M.F was supported by the Air Force Office of Scientific Research under Award No. FA9550-21-1-0423.  D.S.A. and R.M.F acknowledge the hospitality of KITP, where part of this work was done. KITP is supported in part by the National Science Foundation under Grant No. NSF PHY-1748958. 
\end{acknowledgments}

\bibliography{main-mirror-chern-bands-altermagnets}

\end{document}


\title{Supplemental Material for \\  ``Mirror Chern Bands and Weyl Nodal Loops in Altermagnets''}

\author{Daniil S. Antonenko}
\affiliation{Department of Physics, Yale University, New Haven, Connecticut 06520, USA}

\author{Rafael M. Fernandes}
\affiliation{School of Physics and Astronomy, University of Minnesota, Minneapolis, Minnesota 55455, USA}

\author{J{\"o}rn W. F. Venderbos}
\affiliation{Department of Physics, Drexel University, Philadelphia, Pennsylvania 19104, USA}
\affiliation{Department of Materials Science \& Engineering, Drexel University, Philadelphia, Pennsylvania 19104, USA}

\date{\today}

\maketitle

\tableofcontents

\section{Symmetry properties of the models and Ginzburg-Landau theory \label{sec:GL}}

In the main text, we introduced two models for two-dimensional (2D) and three-dimensional (3D) altermagnetism. In this Supplemental Section, we compare their symmetry properties and use these symmetry properties to derive the Ginzburg-Landau (GL) free energies for the localized magnetic moments. Recall that in the main text we focused on the properties of itinerant electrons assuming the configuration of localized moments to be frozen in an anti-aligned order with staggered magnetization (dubbed ``N\'eel vector'' hereafter) $\bm{N}$ and coupled to electrons via an onsite Kondo term. As explained in the main text, the direction of the magnetic moments becomes crucial because it affects the behavior of electrons once spin-orbit-coupling is introduced, which also pins $\bm{N}$ to certain lattice directions.
So, to find out which local-spin configurations should be considered, we determine here which directions of $\bm{N}$ minimize the Ginzburg-Landau free energy. 

\subsection{Comparison of the symmetry properties of 2D and 3D models}
\label{sec:models-comparison}

Both 2D and 3D models have the same point group, $D_{4h}$. However, the 2D model is symmetric with respect to regular group operations, while for the 3D rutile-type lattice the group includes non-symmorphic transformations. For the 2D lattice, shown in Fig. 1(a) of the main text, the origin of the coordinate system is located at a non-magnetic site. It follows then that the sublattices $A$ and $B$ are exchanged either by a $C_{4z}$ rotation or by a  $C_{2x^\prime}$ rotation with respect to the diagonal in the $Oxy$ plane. In contrast, the sublattices remain invariant under a two-fold rotation with respect to the $x$-axis, $C_{2x}$ . Note that all rotation axes contain the origin. For the 3D lattice of Fig. 1(b) of the main text, the origin is at one of the magnetic sites. This choice corresponds to the standard setting of the $P4_2/mnm$ (\#136) space group of the rutile lattice. In this setting, the diagonal two-fold rotation  $C_{2x^\prime}$  is still a symmetry of the lattice, which in this case leaves the sublattices invariant, however (as before, all rotation axes cross the origin).  On the other hand,  $C_{4z}$ and $C_{2x}$ rotations are not symmetry operations: while they map the magnetic sites onto a magnetic site, the non-magnetic sites are not mapped onto themselves. This can be circumvented if the rotations are complemented with a (half) translation by the  lattice vector $(1/2,1/2,1/2)$, which is indicated as $\{ \mathcal C_{4z}|\frac12\frac12\frac12 \}$ and $\{ \mathcal C_{2x}|\frac12\frac12\frac12 \}$. Now, because of this additional half-translation imposed by the non-magnetic sites, the two magnetic sublattices are exchanged under these non-symmorphic operations. 

The difference in the realization of symmetry operations has important consequences. To show this, we classify the Hamiltonian constituents into irreducible representations (irreps) of the $D_{4h}$ point group by inspecting the action of each element of the point group and summarizing the results in Table~\ref{tbl:irreps}. Aiming at deriving a Ginzburg-Landau theory and a tight-biding Hamiltonian in the presence of spin-orbit coupling, we consider here symmetry operations that act jointly in the lattice and spin spaces. 

We start by classifying how the Pauli matrices in the sublattice space ($\tau_a$) transform in the 2D and 3D lattices under the action of point group elements. Each element either preserves sublattices and, hence, acts trivially on $\tau_i$, or exchanges A and B sites. Such sublattice exchange can be represented as the operator $U_{\text{exch}} = \tau_x$ acting on the Bloch states in the sublattice space: $| \psi \rangle \rightarrow U_{\text{exch}} | \psi \rangle$. According to general linear algebra, the corresponding transformation law for operators is $\mathcal{O} \rightarrow U_{\text{exch}}^{\dagger} \mathcal{O} U_{\text{exch}}$. Substituting $\tau_a$ for $\mathcal{O}$, we see that $\tau_x$ and $\tau_0$ are always preserved while $\tau_y$ and $\tau_z$ flip sign under sublattice exchange. Therefore, the $\tau_{x,0}$ matrices are invariant under all point group operations and belong to the trivial $A_{1g}$ irrep, while the irrep of $\tau_{y,z}$ depends on the model. In the 2D model,  because $\mathcal{C}_{4z}$ and $\mathcal{C}_{2x^\prime}$ exchange the sublattices, $\tau_{y,z}$ are odd under these two operations, but even under $C_{2x}$. The irrep of $D_{4h}$ that satisfies these properties is $B_{1g}$. The same analysis in the 3D lattice reveals that  $\tau_z$ and $\tau_y$ transform as the $B_{2g}$ irrep, since in this case they are odd under the operations $\{ \mathcal C_{4z}|\frac12\frac12\frac12 \}$ and $\{ \mathcal C_{2x}|\frac12\frac12\frac12 \}$, but even under  $\{ \mathcal C_{2x'}|0 0 0 \}$ (Table~\ref{tbl:irreps}).

As for the net magnetization $\bM$ and the electron spin $\bsigma$, they have a trivial sublattice structure. Therefore, in both 2D and 3D lattices, they have the same classification: $M_z$ and $\sigma_z$ transform as $A_{2g}$ while $\{M_x, M_y\}$ and $\{\sigma_x, \sigma_y\}$ belong to the two-dimensional irrep $E_g$.  The situation is different for the N\'eel vector $\bN = \bM_A - \bM_B$ since, under sublattice exchange, it transforms into $\bM_B - \bM_A = -\bN$. Thus, compared to $\bM$, the N\'eel vector $\bN$ gets an additional sign flip under those symmetry operations that exchange sublattices. This means that $\bN$ has the same symmetry properties as $\tau_z \bM$. Therefore, the irrep of $\bN$ is the product of the irreps corresponding to $\bM$ and $\tau_z$. Using the results from the paragraphs above, we conclude that the out-of-plane ($\parallel \hat{\bz}$) component, $N_z$, belongs to the $A_{2g} \times B_{1g} = B_{2g}$ irrep in the 2D model ($d_{xy}$ symmetry), while in the 3D case it belongs to the $A_{2g} \times B_{2g} = B_{1g}$ ($d_{x^2 - y^2}$ symmetry) irrep. Therefore, for the case of an out-of-plane  N\'eel vector, the general altermagnetic features of the two models, such as the direction of the spin-splitting nodal lines, are mapped onto each other by a  $45^\circ$-rotation around $Oz$. 
This $45^\circ$-mapping can be seen in the Hamiltonians, upon comparing the $t_d$ terms in Eqs.~(1) and (4) of the main text and the spin-orbit coupling terms in Eqs. (3) and (5), first line. Thus, the two models considered in this paper are minimal representatives of altermagnets with $d_{xy}$ and $d_{x^2 - y^2}$ spin splitting on the square (tetragonal) lattice.

To determine the irrep of the in-plane ($\perp \hat{\bz}$) components of the N\'eel vector, one can use the method of the previous paragraph and multiply the irreps of $\{M_x, M_y\}$ and $\tau_z$, which gives the same result for both the 2D and 3D models: $E_g \times B_{1g} \simeq E_g \times B_{2g} \simeq E_g$. However, the irreps are identical in different bases. It is therefore convenient to express the components $N_x$ and $N_y$ in the same bases as $\{M_x, M_y\}$ . We find that the action of the point group is the same on the doublets $\{M_x, M_y\}$ and $\{N_x, -N_y\}$ in the 2D model, compared to $\{M_x, M_y\}$ and $\{N_y, N_x\}$ in the 3D model. This subtlety will be important when deriving the symmetry-allowed coupling between $\bM$ and $\bN$ in the presence of SOC.
\begin{table}[h]
\caption{Classification with respect to the irreducible representations (irreps) of the $D_{4h}$ point group for the 2D and 3D models of the paper.} \label{tbl:irreps}
\begin{ruledtabular}
\begin{tabular}{cccccc}
& $A_{1g}$ ($id$) & $A_{2g}$ & $B_{1g}$ & $B_{2g}$ & $E_g$ \\
\hline 
2D model & $\tau_0$, $\tau_x$ & $M_z$, $\sigma_z$ & $\tau_z$, $\tau_y$, $d_{x^2 - y^2}$ & $N_z$, $d_{xy}$ & $\{M_x, M_y\}$, $\{\sigma_x, \sigma_y\}$, $\{N_x, -N_y\}$ \\
3D model & $\tau_0$, $\tau_x$ & $M_z$, $\sigma_z$ & $N_z$, $d_{x^2 - y^2}$ & $\tau_z$, $\tau_y$, $d_{xy}$ & $\{M_x, M_y\}$, $\{\sigma_x, \sigma_y\}$, $\{N_y, N_x\}$
\end{tabular}
\end{ruledtabular}
\end{table}

\subsection{Ginzburg-Landau free energy \label{ssec:in-plane-Neel}}

In the presence of SOC, the out-of-plane ($\parallel \hat{\bm{z}}$) and in-plane ($\perp \hat{\bm{z}}$) components of the N\'eel vector belong to different irreducible representations and thus should be treated independently. Which of them corresponds to the leading instability is determined by microscopics and is not predicted by Ginzburg-Landau (GL) theory. Therefore, in the present work we consider both possibilities. We employ the irrep classification of the variables listed in Table~\ref{tbl:irreps}  and  demand that the free energy transforms trivially (as $A_{1g}$) under the action of the point group.

Consider first the case where the N\'eel vector points along the $z$ direction. Since the order parameter is a scalar, the GL free energy takes the simple form $F = a  N^2_x + u N^4_z$ with a well-known description of the transition. No coupling between $N_z$ and any components of the magnetization $\bM$ is allowed as can be inferred from the representation theory (see Table~\ref{tbl:irreps}).

Consider next the case of an in-plane N\'eel vector. As shown in Section \ref{sec:models-comparison},
the in-plane components form a two-dimensional irreducible representation $E_g$ both in the 2D and 3D models, so the same GL theory for a two-component order parameter with $E_{g}$ symmetry is applicable. 
Recall, however, that while the order parameters have the same symmetry in the two models, they transform differently under the symmetries that exchange sublattices (see the last column of Table~\ref{tbl:irreps}). For instance, $N_x$ is even under the twofold rotation $C_{2x}$ in the 2D model, but is odd under the same rotation in the 3D model.
This will be reflected in the form of the allowed coupling to the magnetization $\bM$. 

To determine the possible ground states for the two-component order parameter $(N_x, N_y)$, we write down the GL free energy for an $E_g$ order parameter:
\be
F = a  (N^2_x+N^2_y)+u_1 (N^4_x+N^4_y) +2u_2 N^2_xN^2_y + F_{N,M} \label{GL-Nxy}
\ee
Note that there are two symmetry-allowed quartic terms, which determine the structure of the ground state below the N\'eel temperature $T_N$ , as well as the $ F_{N,M}$ term, which contains the allowed coupling to the magnetization $\bM$. The form of $ F_{N,M} $ depends on the model and will be considered below.

We first bring $F$ to the form discussed in the main text. We introduce the polar decomposition $(N_x, N_y)= N(\cos\theta, \sin\theta)$ and after substitution we obtain
\be
F =  a_N  N^2 +u_N N^4  +\gamma_N N^4 \cos 4\theta + F_{N,M}.  \label{GL-Nxy-polar}
\ee
The GL expansion coefficients $u_N$ and $\gamma_N$ are related to $u_{1,2}$ via
\be
u_N = \frac14( 3 u_1 + u_2), \qquad \gamma_n = \frac14 (u_1-u_2).
\ee
As is clear from \eqref{GL-Nxy-polar}, the internal structure of the order parameter below $T_N$ is determined by the sign of $\gamma_N$. In particular, when $\gamma_N<0$ the optimal angle $\theta$ is given by $\theta=n\pi/2$ ($n=0,1,2,3$). This corresponds to a N\'eel vector along the principal axes, i.e., $\hat x$ or $\hat y$. Instead, when $\gamma_N>0$ the optimal angle $\theta$ is given by $\theta=\pi/4+n\pi/2$ ($n=0,1,2,3$), which corresponds to a N\'eel vector along the in-plane diagonals. In the former case, the system retains an in-plane twofold rotation symmetry about the direction of the N\'eel vector, whereas in the latter case the system retains an in-plane twofold rotation symmetry about an axis perpendicular to the N\'eel vector. Since such a perpendicular twofold rotation is also about a diagonal, it exchanges the sublattices and therefore implies that the moments on the two sublattices have equal size.  

Consider now the coupling to the magnetization components $(M_x,M_y)$ given by $ F_{N,M} $. Since both $\{M_x, M_y\}$ and $\{N_x, N_y\}$ transform as $E_g$ in both models, we should consider the product of the representations $E_g \otimes E_g = A_{1g} \oplus A_{2g} \oplus B_{1g} \oplus B_{2g}$. It contains the trivial irreducible representation $A_{1g}$, which means that a certain combination of the pairwise component products $N_i M_j$ is invariant under all group operations and should be added to the GL functional. Crucially, the group elements act differently on $\{M_x, M_y\}$ and $\{N_x, N_y\}$ despite they belonging to the same irreducible representation. 
We discussed this peculiarity in the end of Section \ref{sec:models-comparison} and wrote the in-plane components of $\bM$ and $\bN$ in the same basis in the last column of Table~\ref{tbl:irreps}. Using this result, we obtain the explicit expression for the invariant combination for the 2D model:
\be
F_{N,M}^{\text{2D}} = \kappa (N_x M_x - N_y M_y). \label{F_NM-2D}
\ee
Thus, when the N\'eel vector points along a principal axis (e.g., $\bN \propto \hat x$), the induced magnetization is parallel to the the N\'eel vector. Conversely,  when the N\'eel vector points along a body diagonal (e.g., $\bN \propto \hat x + \hat y$), the induced magnetization is perpendicular to the the N\'eel vector. The former case gives rise to an induced ferrimagnetic configuration, with collinear moments of unequal length on the two sublattices. The latter induces a canted (i.e. non-collinear) configuration.

Repeating the same steps as above, in the 3D model, the coupling between the N\'eel vector and the magnetization takes the form
\be
F_{N,M}^{\text{3D}} = \kappa (N_x M_y + N_y M_x).  \label{F_NM-3D}
\ee
The situation is reversed compared to \eqref{F_NM-2D}, as the resulting state is a canted configuration when the N\'eel vector points along a principal axis and a ferrimagnetic configuration when it points along the diagonals.

\section{Altermagnetism on the 2D Lieb lattice without SOC \label{sec:2Dmodel}}

In the main text we introduced a tight-binding model for altermagnetism in 2D on the Lieb lattice. In this Supplemental Section we provide additional details of the construction and analysis of this model in the absence of SOC. As explained in the main text, we consider a two-band model for the two magnetic sites of the Lieb lattice sitting at positions $(0,\frac12)$ and $(\frac12,0)$. In the final part of this section we describe how our two-band model can be obtained from a three-band model by projecting out the nonmagnetic site at position $(0,0)$. 

The general form of the Hamiltonian is defined as
\be
H = \sum_{\bk} c^\dagger_{\bk } \mathcal H(\bk) c_{\bk}, \qquad c^\dagger_{\bk} = \begin{pmatrix} c^\dagger_{ \bk A \up}  & c^\dagger_{\bk B \up}&  c^\dagger_{ \bk A \down}  & c^\dagger_{\bk B \down}\end{pmatrix},  \label{H-def}
\ee
and, without SOC, $\mathcal H$ takes the form (suppressing momentum dependence)
\be
\mathcal H = \mathcal H_t + \mathcal H_{\text{AM}} \equiv \mathcal H_0.  \label{H-2D}
\ee
Here $\mathcal H_t$ simply describes the hopping of electrons and $ \mathcal H_{\text{AM}}$ describes the (alter)magnetism. To capture the essence of altermagnetism, it is important that $\mathcal H_t$ reflects the true symmetry of the crystal lattice and thus includes symmetry-allowed hopping anisotropies. 

\subsection{Construction of the Hamiltonian \label{ssec:ham}}

Our aim is to construct a generic Hamiltonian compatible with all symmetries of the Lieb lattice in the presence of collinear N\'eel order. By generic we mean a model that does not exhibit non-essential spectral features, i.e., features that can be removed by adding symmetry-allowed couplings in the Hamiltonian. The most important example of essential/non-essential features are spectral degeneracies: a generic model only exhibits essential degeneracies originating from physical symmetries. Two approaches can be used to construct such a Hamiltonian. One may work directly in momentum space by finding all symmetry-allowed combinations of Pauli matrices and momentum-dependent form factors using the classification of Table~\ref{tbl:irreps}, where the form factors represent symmetrized Fourier transformed hoppings. The symmetry operations include point group transformations and time reversal. Altermagnetic ordering, as any other form of magnetism, breaks time-reversal symmetry, $\mathcal{T}$, but the full Hamiltonian is still invariant when time reversal is accompanied by a sign flip of the N\'eel vector: $\mathcal{H} = \mathcal{T}  \mathcal{H}_{\bN \rightarrow - \bN} \mathcal{T}$.
The second approach takes a real-space perspective and is more physically motivated. To make the derivation more informative, we choose the second approach and determine $\mathcal{H}$ by considering real space hopping processes.

To obtain a generic hopping Hamiltonian on the (two-site) Lieb lattice, we include nearest neighbor (1NN) and anisotropic second-nearest neighbor (2NN) hopping. The former is denoted $t_1$ and the latter are denoted $t_{2a} = t_2+t_d$ and $t_{2b}=t_2-t_d$. Using Pauli matrices $\btau$ for sublattice space, this gives
\be
\mathcal H_t(\bk) = -4 t_1 \cos \frac{k_x}{2}  \cos \frac{k_y}{2} \tau_x - 2 t_2 (\cos k_x + \cos k_y)\tau_0 - 2 t_d (\cos k_x - \cos k_y)\tau_z. \label{H_t-2D}
\ee
For future reference it is useful to define the form factor functions
\be
s_\bk \equiv 4  \cos \frac{k_x}{2}  \cos \frac{k_y}{2} , \qquad s'_\bk\equiv 2(\cos k_x + \cos k_y), \qquad d_\bk \equiv2(\cos k_x - \cos k_y). \label{form-factors-2D}
\ee
In terms of these form factor functions, the Hamiltonian can be compactly written as $\mathcal H_t(\bk) = - t_1 s_\bk \tau_x -  t_2 s'_\bk \tau_0 -  t_d d_\bk \tau_z$.

Let us now include collinear intra-unit-cell staggered magnetic order with N\'eel vector $\bN$. This gives rise to an additional term in the Hamiltonian given by
\be
 \mathcal H_{\text{AM}}(\bk)  = J\tau_z  \bN \cdot  \bsigma + J' (\cos k_x - \cos k_y)  \bN \cdot  \bsigma. \label{H_AM-2D}
\ee
Here the first term (with coupling $J$) is the regular Kondo-like interaction with the collinear $Q=0$ N\'eel order on the two sublattices and the second term (with coupling $J'$) is a magnetic octupole term. The presence of a magnetic octupole term reflects a key aspect of altermagnetism, in that the staggered magnetization linearly couples to a higher-order magnetic multipole moment \cite{Fernandes-Altermagnetism}, which in this case is a magnetic octupole  \cite{Spaldin2022,McClarty-AltermagnetismGL}. Indeed, the transformation properties of $\tau_z$ and $\cos k_x - \cos k_y$ listed in Table~\ref{tbl:irreps} confirm that both terms in \eqref{H_AM-2D} have the same symmetry. The presence of the magnetic octupole term furthermore highlights the characteristic momentum-dependent nonrelativistic spin splitting of altermagnets, showing that the two phenomena are inextricably linked (see also \cite{Spaldin2022,Smejkal-AltermagnetismReview,Fernandes-Altermagnetism}).

Although the $J'$ term is symmetry-allowed, it does not give rise to any qualitative features that are not already described by the $J$ term in combination with $\mathcal H_t(\bk) $. In particular, the $J$ term in combination with $\mathcal H_t(\bk) $ already gives rise to nonrelativistic spin splitting due to the presence of $\tau_z$. Therefore, since the $J'$ term is in that sense non-essential, we will set $J'=0$ in all subsequent analysis. Finally, because the N\'eel vector $\bN$ can point in any direction in the absence SOC , the band dispersion is independent of the direction of $\bN$. We may therefore set $\bN = N\hat z$.

\subsection{Symmetries of the Hamiltonian  \label{ssec:sym-2D}}

We now turn to a detailed analysis of the symmetry-enforced energy dispersion degeneracies in the limit of no SOC. We focus on three features in particular: (i) symmetry-enforced degeneracy at the $M$ point, (ii) mirror (or twofold rotation) symmetry eigenvalues on the Brillouin zone (BZ) boundary, and (iii) inversion $\mathcal{I}$ (or twofold $C_{2z}$) eigenvalues at inversion-symmetric high-symmetry points. 

In the analysis of symmetries and their representations in the space of Bloch states, we make use of the gauge matrix $(V_\bG)_{\alpha\beta} = e^{i \bG \cdot \bfell_\alpha}\delta_{\alpha\beta}$, where $\bG = n_1\bfg_1 + n_2\bfg_2$ is a vector of the reciprocal lattice with spanning vectors $\bfg_{1,2}$. The reciprocal lattice vectors $\bfg_{1,2}$ are defined via the condition $\bfg_i \cdot \ba_j = 2\pi \delta_{ij}$, where $\ba_j $ are the real-space lattice vectors. The vectors $\bfell_A = (\frac12, 0)$ and $\bfell_B = (0,\frac12)$ denote the positions of the $A$ and $B$ sublattice sites. 

\begin{figure}
	\includegraphics[width=0.45\textwidth]{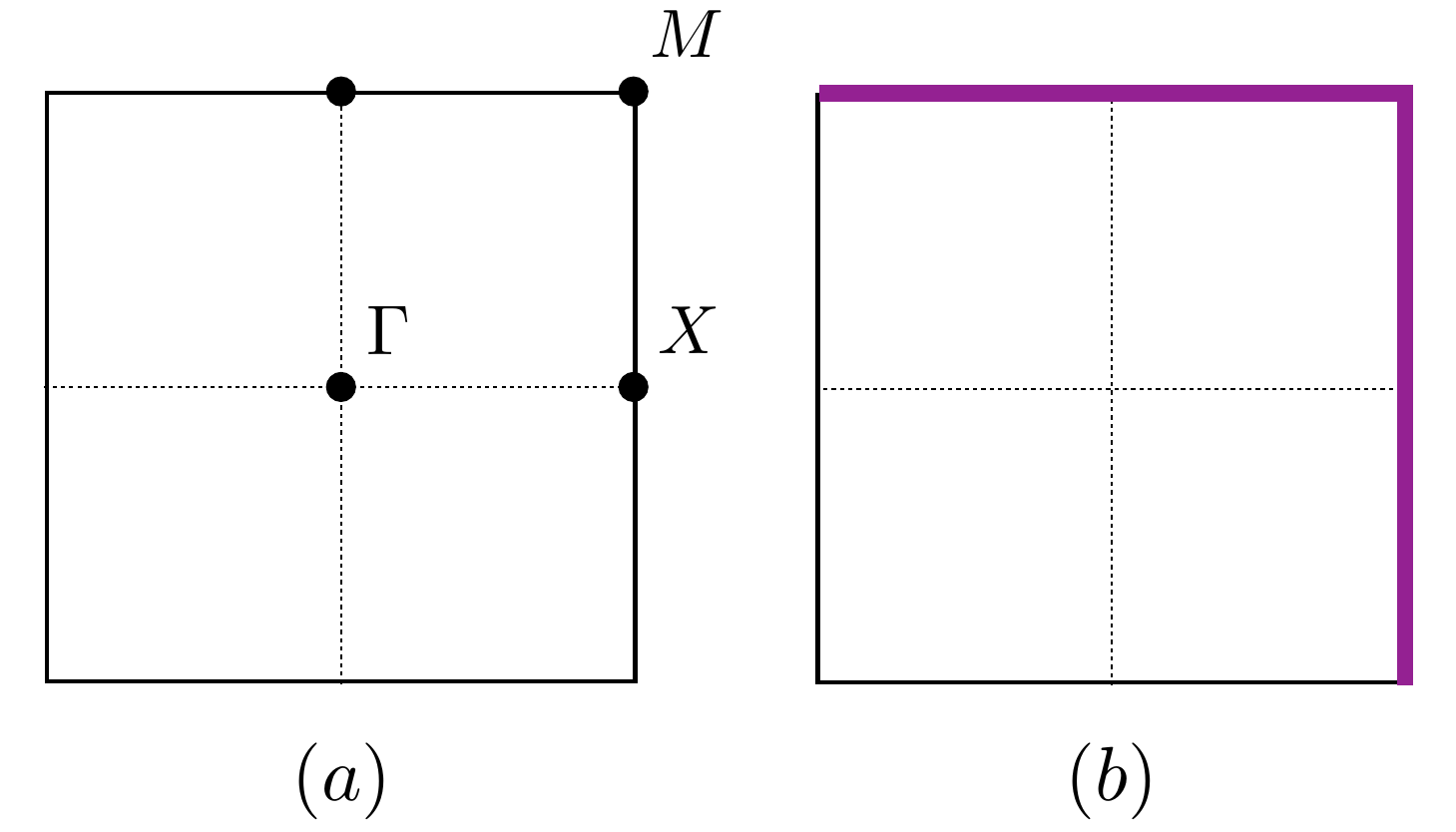}
	\caption{Points and lines of high symmetry. (a) High symmetry points $\Gamma$, $M$, and $X$, corresponding to the wave vectors  $\bQ_\Gamma=(0,0)$,  $\bQ_M=(\pi,\pi)$, and  $\bQ_X=(\pi,0)$. (b) Lines of high symmetry on the Brillouin zone boundary.}
	\label{fig:sym-2D}
\end{figure}

\begin{itemize}

\item {\it The BZ corner $M$.} 

We begin with the $M$ point, which has momentum $\bQ_M=(\pi,\pi)$, and consider the action of the three generators of $D_{4h}$. Note that we have
\be
\mathcal C_{4z}\bQ_M = \bQ_M-\bfg_1 , \qquad \mathcal  C_{2x}\bQ_M = \bQ_M-\bfg_2, \qquad \mathcal  I\bQ_M = \bQ_M-\bfg_1-\bfg_2.
\ee
In terms of the gauge matrix and sublattice operators, this implies that the symmetry operators at the $M$ point take the form
\be
U_{\mathcal  C_{4z}} = V_{-\bfg_1} \tau_x = -\tau_z\tau_x =-i\tau_y, \qquad U_{\mathcal  C_{2x}} = V_{-\bfg_2} = \tau_z,\qquad U_{\mathcal  I} =V_{-\bfg_1-\bfg_2} = -\mathbb{1}. 
\ee
Because $U_{C_{4z}}$  and $U_{C_{2x}}$ do not commute, the states at $M$ must form a two-dimensional irreducible representation, and therefore must be degenerate. In fact, we can conclude from this that the two states at $M$ have $p$-wave symmetry, transform as the representation $E_u$, and must indeed have the same energy. Note that this is different at $\Gamma$, where the bands may split, since they have $A_{1g}$ and $B_{1g}$ symmetry at $\Gamma$. This symmetry-enforced twofold degeneracy is what defines the symmetry protected quadratic band crossing point at $M$ in the absence of magnetism~\cite{FradkinKivelson}.

\item {\it The BZ boundaries $MX$ and $MY$.}

Now let us consider the BZ boundary defined by $\bQ_{MX}=(\pi,q)$. This boundary is invariant under the mirror symmetry $\mathcal  M_x = \mathcal I \mathcal C_{2x}$ and we can therefore label all energy bands by their mirror eigenvalues. First observe that 
\be
\mathcal M_x \bQ_{MX} = \bQ_{MX}- \bfg_1,
\ee
which means that the mirror symmetry operator in the space of Bloch states is
\be
U_{\mathcal M_x} = V_{-\bfg_1} = \begin{pmatrix} -1 & 0 \\ 0  &1  \end{pmatrix}.
\ee

From this we conclude that states arising from the two sublattices have opposite mirror eigenvalues, and thus cannot mix on the BZ boundary. A similar result obviously holds for the boundary defined by $\bQ_{MY}=(q,\pi)$, for which one finds the mirror $\mathcal M_y$ operator
\be
U_{\mathcal M_y} = V_{-\bfg_2} = \begin{pmatrix} 1 & 0 \\ 0  &- 1  \end{pmatrix}.
\ee

\item {\it Representations of $\mathcal C_{2z}$ and $\mathcal I$ at invariant momenta.}

It is also useful to determine the symmetry representations of $\mathcal C_{2z}$ and $\mathcal I$ at the invariant momenta $\Gamma$, $X$, $Y$, and $M$. The reason is that the eigenvalues of these symmetries at the invariant momenta can be used to determine whether a gap closing transition is topological. Note that, in the absence of SOC, $\mathcal C_{2z}$ and $\mathcal I$ are equivalent in our 2D model and we only need to consider one of them; we choose $\mathcal C_{2z}$. 
Importantly, $\mathcal  C_{2z}$ does not exchange sublattices. At the $\Gamma$ point, we clearly have $U^\Gamma_{\mathcal C_{2z}}   = \mathbb{1}$. Since $\mathcal C_{2z}$ acts on $M$ as
\be
\mathcal C_{2z}  \bQ_{M} =  \bQ_{M} - \bfg_1 -\bfg_2,
\ee
we therefore have
\be
U^M_{\mathcal C_{2z}}  = V_{-\bfg_1-\bfg_2} = -  \mathbb{1}. 
\ee
The representations of $\mathcal C_{2z}$ at $X$ and $Y$ immediately follow from our consideration of the BZ boundaries. We have
\be
U^X_{\mathcal C_{2z}} = V_{-\bfg_1} = \begin{pmatrix} -1 & 0 \\ 0  &1  \end{pmatrix}, \qquad U^Y_{\mathcal C_{2z}} = V_{-\bfg_2} = \begin{pmatrix} 1 & 0 \\ 0  & -1  \end{pmatrix}.
\ee
We conclude that at $\Gamma$, the two bands arising from different sublattices have $\mathcal C_{2z}$ and $\mathcal{I}$ eigenvalue $+1$, whereas at $M$, both bands have eigenvalue $-1$. Furthermore, the Hamiltonian must be diagonal at both $X$ and $Y$ if $\mathcal C_{2z}$ and $\mathcal{I}$ are preserved. The two bands have opposite eigenvalues and their ordering will depend on their energies.

\end{itemize}

\subsection{Analysis of the Hamiltonian and the spectrum  \label{ssec:ham-analysis-2D}}

In this section we provide a detailed analysis of the Hamiltonian $\mathcal H_0$ and its energy spectrum. As mentioned above, we may take the N\'eel vector to be along $\hat z$, i.e., $\bN = N\hat z$, which yields
\be
\mathcal H^{\up,\down}_0(\bk) = \begin{pmatrix} \pm JN -t_2 s'_\bk - t_d d_\bk  & -t_1 s_\bk  \\ -t_1 s_\bk  & \mp JN -t_2 s'_\bk + t_d d_\bk  \end{pmatrix} =  -t_2 s'_\bk \tau_0 -t_1 s_\bk \tau_x -(t_d d_\bk \mp JN)\tau_z , \label{ham-2D-noSOC}
\ee
It is straightforward to obtain the full energy spectrum; one finds
\be
\mathcal E^{\sigma}_\pm(\bk) =  -t_2 s'_\bk  \pm \sqrt{t^2_1 s^2_\bk + (t_d d_\bk -\sigma JN)^2}.
\ee
It is clear that a spin splitting can only occur when $t_d$ is nonzero, and furthermore that the spin splitting vanishing on the lines in momentum space for which $d_\bk = 0$.

We now discuss specific features of the Hamiltonian by focusing on lines or points of high symmetry. 

\begin{itemize}

\item {\it Dirac points on the Brillouin zone boundaries, $MX$ and $MY$.}

Consider first the Brillouin zone boundary $MX$ defined by $k_x=\pi$. This line is invariant under the mirror symmetry $\mathcal M_x$ and, as shown in Sec.~\ref{ssec:sym-2D}, this implies that the Hamiltonian commutes with $U_{\mathcal M_x}  = -\tau_z$ and is thus diagonal. Setting $k_x=\pi$ we find that \eqref{ham-2D-noSOC} reduces to
\be
\mathcal H^{\up,\down}_0(k_y) = \begin{pmatrix} \pm JN -t_2 s'_{\pi,k_y} - t_d d_{\pi,k_y}  & 0  \\0 & \mp JN -t_2 s'_{\pi,k_y} + t_d d_{\pi,k_y}  \end{pmatrix} ,
\ee
which is consistent with the requirement that it commutes with $U_{\mathcal M_x} $. Simplifying the form factors for $k_x=\pi$ yields,
\be
s'_{\pi,k_y} = -2 + 2\cos k_y = -4\sin^2(k_y/2), \qquad d_{\pi,k_y} = -2 - 2\cos k_y = -4\cos^2(k_y/2),
\ee
such that the Hamiltonian can be rewritten as
\be
\mathcal H^{\sigma}_0(k_y) = 4 t_2 \sin^2\frac{k_y}{2} \tau_0 +  \left(  4t_d \cos ^2\frac{k_y}{2}+\sigma  JN  \right)  \tau_z. 
\ee
This form shows directly that band crossings exist whenever
\be
4t_d \cos ^2\frac{k_y}{2}= -\sigma  JN ,
\ee
which is the case as long as
\be
0< -\sigma  JN/4t_d <1. \label{2D-Dirac-condition}
\ee
This implies that crossings can exist on the $k_x=\pi$ only in one spin sector, which depends on the sign of $JN/t_d$.

The situation on the $k_y=\pi$ line is similar, except that band crossings, when they exist, occur in the opposite spin sector. 

\item {\it Quadratic band crossing (QBC) point at $M$.}

Next we consider the special high symmetry momentum $\bQ_M=(\pi,\pi)$ and expand the Hamiltonian in small momentum $\bq$ around this point. The lattice harmonics are expanded as 
\be
s'_{\bQ_M+\bq } =  q^2_x+q^2_y-4= q^2-4, \qquad  s_{\bQ_M+\bq } = q_x q_y,\qquad d_{\bQ_M+\bq } = q^2_x-q^2_y,
\ee
and this leads to a continuum Hamiltonian given by
\be
\mathcal H^{\sigma}_0(\bQ_M+\bq) = t_2(4 -q^2 )\tau_0 + [\sigma JN -t_d(q^2_x - q^2_y) ]\tau_z -t_1 q_xq_y \tau_x  . \label{QBC}
\ee
This Hamiltonian can be recognized as the Hamiltonian of a quadratic band crossing (QBC) in the presence of a symmetry breaking term $\sigma JN\tau_z$, as first considered in detail in Ref.~\onlinecite{FradkinKivelson}. As pointed out in Ref.~\onlinecite{FradkinKivelson}, and shown also in Sec.~\ref{ssec:sym-2D}, when $JN=0$ the QBC is symmetry protected. A nonzero $JN $ splits the QBC into two Dirac points in each spin sector. Consider for concreteness the case $JN>0$. Then the condition
\be
\sigma JN -t_d(q^2_x - q^2_y)  = 0
\ee
implies that in the spin-$\up$ sector Dirac points exist on the $q_x$ axis, whereas in the spin-$\down$ sector Dirac points exist on the $q_y$ axis. Hence, the QBC point is split in orthogonal directions for the two spin projections, in agreement with the $d$-wave nature of the altermagnet.

\item {\it Merger and removal of Dirac points at $X$ and $Y$.}

Next, consider an expansion of the Hamiltonian around $\bQ_X=(\pi,0)$; the situation at $\bQ_Y=(0,\pi)$ is analogous. The lattice harmonics are expanded as
\be
s'_{\bQ_X+\bq } = q^2_x-q^2_y, \qquad  s_{\bQ_X+\bq } = -2 q_x ,\qquad d_{\bQ_X+\bq } =  q^2_x+q^2_y-4,
\ee
and this leads to a continuum Hamiltonian around $\bQ_X=(\pi,0)$ given by
\be \label{eq:X_exp_no_SOC}
\mathcal H^{\sigma}(\bQ_X+\bq) = -t_2(q^2_x-q^2_y )\mathbb{1} + [t_d(4-q^2) +\sigma J N]\tau_z + t_1 2 q_x \tau_x .
\ee
Consistent with what was found above, when $q_x=0$ the Hamiltonian is diagonal. Importantly, it is gapless for one of the spin species at $X$ when $|JN/t_d| = 4$, which, according to \eqref{2D-Dirac-condition}, marks the transition from a Dirac nodal phase to a gapped phase. Let us assume that $JN, t_d>0$. Then the gapless point emerges only in the spin-$\down$ sector. At this gapless point, the Hamiltonian becomes
\be
\mathcal H^{\down}(\bQ_X+\bq)  = -t_2(q^2_x-q^2_y )\mathbb{1} -t_d q^2 \tau_z + t_1 2 q_x \tau_x , \label{eq:X_exp_no_SOC_gapless}
\ee
which shows that the dispersion is quadratic in $q_y$ and linear in $q_x$.

\end{itemize}

\subsection{Projection from the three-band Lieb lattice model \label{ssec:3-band}}

In this Supplemental section we consider the full three-band Lieb lattice model (see e.g. Ref. \cite{Sudbo-minimal}) and show that our effective two-band model can be obtained by projecting out the third band associated with the nonmagnetic site.

The three-band Lieb lattice model is constructed by explicitly including the third nonmagnetic site sitting at position $(0,0)$. Recall that magnetic sites (labeled $A$ and $B$) sit at $(\frac12,0)$ and $(0,\frac12)$. We label the nonmagnetic site $C$ and construct a Hamiltonian of the general form
\be
H = \sum_\bk c^\dagger_\bk \mathcal H(\bk) c_\bk, \qquad c_\bk = \begin{pmatrix} c_{\bk A}  \\ c_{\bk B} \\ c_{\bk C} \end{pmatrix},
\ee
where spin indices were suppressed for simplicity. It is important to emphasize that the Lieb lattice model defined in this was has the same symmetries as the two-band model introduced in the main text. The three-band model simply adds an additional band. As a result, all previously considered terms in the Hamiltonian remain allowed, but are supplemented with couplings to the $C$ site. 

A three-band Hamiltonian (excluding magnetism) then takes the form
\be
\mathcal H_t(\bk) = - t_1 s_\bk \begin{pmatrix} \tau_x &  0 \\ 0&  0  \end{pmatrix} -  t_2 s'_\bk\begin{pmatrix}  \tau_0 &  0 \\ 0&  0  \end{pmatrix} -  t_d d_\bk \begin{pmatrix} \tau_z &  0 \\ 0&  0  \end{pmatrix}  + \begin{pmatrix} 0 &  0 \\ 0&  \varepsilon_0  \end{pmatrix}  -2 t_0 \begin{pmatrix} 0 &  D \\ D^\dagger &  0  \end{pmatrix} ,
\ee
where we have introduced an implicit block notation of the Hamiltonian and defined
\be
D = \begin{pmatrix}   \cos \frac{k_x}{2} \\  \cos \frac{k_y}{2}  \end{pmatrix}.
\ee
Therefore, the three-band model includes an isotropic hopping $t_0$ to the $C$ site, as well as an on-site energy $\varepsilon_0$. The latter is allowed since there is no symmetry which relates the $C$ site to the $A$ and $B$ sites. Note that, as far as symmetry is concerned, direct hopping between the $C$ sites is allowed but is ignored here, since our focus is on the regime in which $\varepsilon_0$ is large, thus giving rise to a separation of energy scales.

It is instructive to relate this three-band model to previous Lieb lattice models, in particular recent work highlighting connections to altermagnetism. Early work on Lieb lattice models focused on the appearance of a perfectly flat band sitting at the same energy as a linear Dirac crossing of the two other bands~\cite{Franz2010}. The uniform Lieb lattice limit is achieved when only $t_0$ is nonzero, namely, when there is no direct hopping between $A$ and $B$, and no intra-sublattice hopping. This is, of course, a fine-tuned limit, since there is no physical symmetry that forbids other terms. 

In the context of altermagnetism, a simplified three-band Lieb-lattice model was recently proposed in Ref.~\onlinecite{Sudbo-minimal} by Brekke, Brataas, and Sudbo. The BBS model is given by
\be
\mathcal H(\bk)  = - t_1 s_\bk \begin{pmatrix} \tau_x &  0 \\ 0&  0  \end{pmatrix}+ \begin{pmatrix} 0 &  0 \\ 0&  \varepsilon_0  \end{pmatrix}  -2 t_0 \begin{pmatrix} 0 &  D \\ D^\dagger &  0  \end{pmatrix} , \label{BBS-model}
\ee
and thus only includes the symmetric hopping between the $A$ and $B$ sublattices ($t_1$), the potential difference between the sites ($\varepsilon_0$), and hopping between the $A,B$ and $C$ sites. Note that in the two-band model we constructed the system knows about the different crystalline environments of the $A$ and $B$ sites through the coupling $t_d$. In the BBS model, the different crystalline environments are encoded in the combination $t_0$ and $\varepsilon_0$. 

It turns out that a version of our two-band model can be obtained from \eqref{BBS-model} by projecting out the nonmagnetic site. To this end, we assume that the nonmagnetic site sits at a higher energy $\varepsilon_0$, which allows us to restrict the Hamiltonian to the two magnetic sites and include coupling to the $C$ site perturbatively. This yields a Hamiltonian of the form
\be
\mathcal H_{2\times 2}(\bk)  = - t_1 s_\bk \tau_x + \delta \mathcal H(\bk)  
\ee
where $\delta \mathcal H(\bk) $ is the correction computed in second-order perturbation theory and given by
\be
\delta \mathcal H(\bk) = -\frac{4 t^2_0}{ \varepsilon_0 } D D^\dagger = -\frac{4 t^2_0}{ \varepsilon_0 }  \begin{pmatrix}   \cos^2 \frac{k_x}{2} &    \cos \frac{k_x}{2}  \cos \frac{k_y}{2} \\   \cos \frac{k_x}{2}  \cos \frac{k_y}{2}&    \cos^2 \frac{k_y}{2}  \end{pmatrix} = -\frac{2 t^2_0}{ \varepsilon_0 } -\frac{ t^2_0}{ \varepsilon_0 } \begin{pmatrix} 2  \cos k_x  &   s_\bk  \\    s_\bk & 2   \cos k_y \end{pmatrix} .
\ee
In addition to an unimportant overall shift of the energy, we find a renormalized first nearest neighbor hopping $-t_1 - t^2_0/\varepsilon_0$, a symmetric second nearest neighbor hopping $t_2 = t^2_0/2\varepsilon_0$, and an anisotropic second nearest neighbor hopping $t_d = t^2_0/2\varepsilon_0$. 

\mbox{}
\subsection{Impact of model parameters on the electronic dispersion \label{sec:t2t1study}}

\begin{figure}[h]
\includegraphics[width=0.9\textwidth]{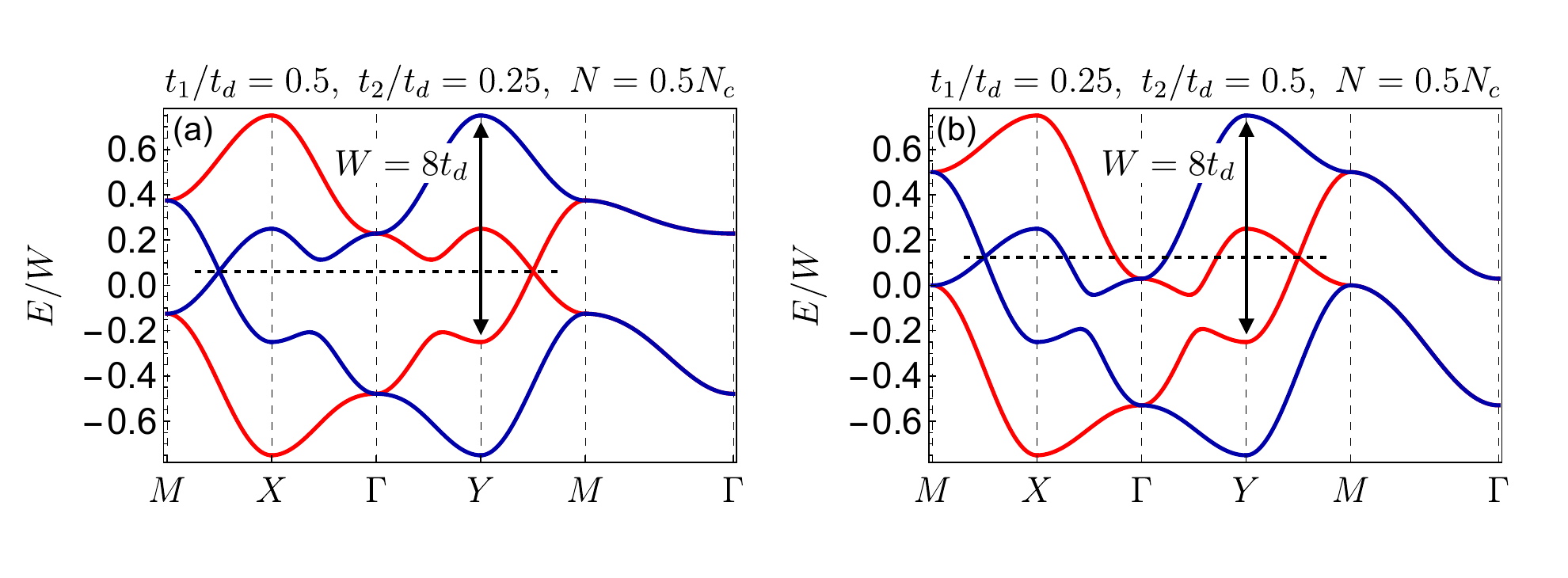}
	\caption{Comparison of the electron dispersion for different 2D model parameters in the absence of spin-orbit coupling. (a) Reproduction of the main text Fig.~2(a) corresponding to $t_1 / t_d = 0.5$, $t_2 / t_d = 0.25$, $N / N_c = 0.5$. (b) Electron dispersion for $t_1 / t_d = 0.25$, $t_1 / t_d = 0.5$, $N / N_c = 0.5$. In this case, the Dirac points lie inside the conduction band at half-filling, which would veil their topological properties.}
	\label{fig:SM-additional-dispersion}
\end{figure}

In the main text, we analyzed the electron dispersion of the Hamiltonian (1) for a particular choice of the tight-binding model parameters $t_1$ (nearest neighbor hopping), $t_2$ (isotropic next-nearest neighbor hopping), $t_d$ (anisotropic next-nearest neighbor hopping between A and B sublattices), and effective on-site Kondo term $J N$, see Fig.~2. As was demonstrated in Fig.~2(b), an increase of the $|J N|$ parameter leads to the migration of the Dirac crossings from the M point towards the X and Y points in the Brillouine zone, where they meet and are removed at $N = N_c = |4 t_d / J|$. 
The choice of the hopping amplitudes in Fig.~(2), $t_1 / t_d = 0.5$, $t_2 / t_d = 0.25$ leads to the energy separation of the Dirac points from the rest of the band dispersion. As a result, opening of the topological gap in the presence of spin-orbit coupling (3) is a true gap leading to topologically quantized transport properties, see Section~\ref{ssec:SOC-Nz-QSHE}. Here we plot the bands for different model parameters with a larger ratio of $|t_2 / t_1| = 2$, see Fig.~\ref{fig:SM-additional-dispersion}(b) in comparison with Fig.~\ref{fig:SM-additional-dispersion}(a), which is a reproduction of Fig.~2(a) of the main text. 
As can been from the right panel, for larger $|t_2/t_1|$ the Dirac points  coincide in energy with other parts of the electron dispersion. That would hinder the opening of a true topological gap and observation of quantized topological effects.  One concludes then that materials with smaller values of $|t_2 / t_1|$ are preferred to realize Chern insulators with quantized QSHE. 

\section{Altermagnetism on the 2D Lieb lattice with SOC \label{sec:2Dmodel-SOC}}

In this Supplemental Section we analyze the Lieb lattice model for 2D altermagnetism in the presence of SOC. In the presence of SOC spin is locked to the lattice, such that spatial and spin degrees of freedom transform jointly under the symmetries of the lattice. This has two important consequences for the structure of the Hamiltonian. First, the Hamiltonian will include spin-orbit-coupling terms such that the full Hamiltonian now takes the form
\be
\mathcal H = \mathcal H_t + \mathcal H_{\text{AM}} + \mathcal H_{\text{SO}}  =  \mathcal H_0 + \mathcal H_{\text{SO}}  ,\label{H-2D-SOC}
\ee
where $ \mathcal H_{\text{SO}} $ describes the SOC terms allowed by all symmetries of the crystal lattice. Second, $\mathcal H_{\text{AM}}$ will need to account for the locking of spin to the lattice, and will therefore be modified.

There is one SOC term that is compatible with full $D_{4h}$ symmetry and is given by~\cite{Franz2010}
\be
\mathcal H_{\text{SO}} (\bk) =   \lambda \sin \frac{k_x}{2}  \sin \frac{k_y}{2}   \tau_y \sigma_z.
\ee
As mentioned in the main text, this SOC term is of Kane-Mele type~\cite{KaneMele-QSHE}. To see why it is symmetry-allowed, note from Table \ref{tbl:irreps} that the form factor function $\sin (k_x/2)  \sin (k_y/2)$ has $d_{xy}$ ($B_{2g}$) symmetry and that $\tau_y$ has $d_{x^2-y^2}$ ($B_{1g}$) symmetry. The latter follows from the fact that $\tau_y$ is odd under all symmetries which exchange sublattices. Their combination thus transforms as the $B_{1g}\times B_{1g} = A_{2g}$ irreducible representation, which corresponds precisely to the symmetry of $\sigma_z$, i.e. the component of spin angular momentum along the $z$ direction.

Now consider the term $\mathcal H_{\text{AM}}$ in the presence of SOC. As described in Sec.~\ref{sec:GL}, we must distinguish $N_z$ and $(N_x,N_y)$, i.e., an out-of-plane N\'eel vector from an in-plane N\'eel vector. In the case of an out-of-plane N\'eel vector, $\mathcal H_{\text{AM}}$ retains the same form as without SOC:
\be
\mathcal H_{\text{AM}} (\bk) =  JN_z \tau_z \sigma_z +2 J' N_z (\cos k_x - \cos k_y) \sigma_z.
\ee
On the other hand, in the case of an in-plane N\'eel vector, $\mathcal H_{\text{AM}}$ acquires an additional term proportional to the magnetization $\bM$, which is induced by the N\'eel order as discussed in Sec. \ref{sec:GL}. The Hamiltonian $\mathcal H_{\text{AM}}$ then becomes
\be
\mathcal H_{\text{AM}} (\bk) =  J \tau_z (N_x \sigma_x + N_y\sigma_y )  +2 J'  (\cos k_x - \cos k_y) (N_x \sigma_x + N_y\sigma_y ) + h (M_x \sigma_x + M_y\sigma_y). \label{H_AM-2D_SOC_inplane}
\ee
Since $\bM$ is determined by $\bN$ per Eq.~\eqref{F_NM-2D}, an alternative way to write the last term is $\lambda_{\text{FM}}  (N_x \sigma_x -N_y\sigma_y)$. As argued in Sec.~\ref{sec:2Dmodel}, we can set $J'=0$.

\subsection{Symmetries in the presence of SOC \label{ssec:sym-SOC}}

We now briefly consider the action of the symmetries in the presence of SOC. Crucially, in the presence of SOC symmetries act jointly on spatial and spin degrees of freedom. The appropriate representations of point group symmetries are double-valued representations. Note further that when SOC is included, the sets of symmetries that are broken and preserved depend on the direction of the N\'eel vector. 

In this brief discussion we focus on two features in particular: (i) the presence of mirror or twofold rotation symmetries, and (ii) the presence of mutually anti-commuting symmetries. 

\begin{itemize}

\item {\it Mirror and twofold rotation symmetries.}

The presence of mirror or twofold rotation symmetries can protect band crossings on lines of high symmetry when the bands involved have opposite mirror or rotation eigenvalues. In this work we are particularly interested in the symmetry protection of the Dirac crossings on the $MX$ and $MY$ lines. The eigenvalues of the mirror $\mathcal M_{z}$ and twofold rotation $\mathcal C_{2z}$ symmetries, for instance, are determined by the algebraic relations
\be
\mathcal M_{z}^2 = \mathcal R, \qquad \mathcal C_{2z}^2 = \mathcal R,
\ee
where $\mathcal R $ is rotation by $2\pi$, which equals $-1$ for spin-$1/2$ fermions. Hence, when one of these symmetries is present and leaves a line of high symmetry invariant, the Hamiltonian block diagonalizes into blocks labeled by the mirror eigenvalues $\pm i$. 

\item {\it Mutually anti-commuting symmetries.} 

When the N\'eel vector is in the $z$ direction, the system retains twofold rotation symmetry about the body diagonals and mirror reflection symmetry in the vertical planes passing through the body diagonals. Specifically, consider the twofold rotation $\mathcal C_{2 \hat n}$ with $\hat n = (1,1)/\sqrt{2}$ and the mirror reflection $\mathcal M_{\hat n_\perp} = \mathcal{I} C_{2 \hat n_\perp}$, where $\hat n_\perp = (1,-1)/\sqrt{2}$. The symmetries leave the high symmetry line $(k,k)$ invariant and have the algebraic relation
\be
\mathcal C_{2 \hat n} \mathcal{M}_{\hat n_\perp} = \mathcal R \mathcal{M}_{\hat n_\perp}\mathcal C_{2 \hat n} .
\ee
This relation implies that these symmetries anticommute in the presence of SOC, which in turn implies that all bands must be at least twofold degenerate. In the context of our model this means that, when the N\'eel vector is in the $z$ direction, the Zeeman splitting nodal lines are preserved when SOC is included.

\end{itemize}

\subsection{Electronic structure for out-of-plane N\'eel vector \label{ssec:SOC-Nz}}

We first address the Hamiltonian and its spectral features when the N\'eel vector is perpendicular to the plane. In this case, the system has mirror symmetry $\mathcal M_z$, which acts as $z \rightarrow -z$. The representation of this mirror symmetry in the space of Bloch states is $U_{\mathcal M_z} = -i \sigma_z$, which indeed commutes with the full Hamiltonian $\mathcal H$. It follows that the mirror eigenvalues $\mp i $ coincide with spin $\up, \down$, such that spin can be used to label the decoupled Hamiltonian blocks. 

The full Hamiltonian can then be written in a similar way as in \eqref{ham-2D-noSOC}, but now including the SOC term:
\be
\mathcal H^{\up,\down}(\bk) =   -t_2 s'_\bk \tau_0 -t_1 s_\bk \tau_x -(t_d d_\bk \mp JN)\tau_z \pm \lambda \sin \frac{k_x}{2}  \sin \frac{k_y}{2}   \tau_y . \label{ham-2D-withSOC}
\ee
Recall that the form factor functions were defined in Sec.~\ref{ssec:ham}. It is now straightforward to perform an analysis of the Hamiltonian and its spectral properties for nonzero $\lambda$. Here, we focus on the expansion around the high symmetry points $M$ and $X$ ($Y$).

\begin{itemize}

\item {\it Gapped Dirac points on the BZ boundaries.}

To see that the Dirac points found in the absence of SOC become gapped it is instructive to revisit the expansion around the momentum $\bQ_M$, see Eq.~\eqref{QBC}. In the presence of SOC the continuum Hamiltonian acquires an additional term
\be
\mathcal H^{\sigma}(\bQ_M+\bq) = t_2(4 -q^2 )\tau_0 + [\sigma JN -t_d(q^2_x - q^2_y) ]\tau_z -t_1 q_xq_y \tau_x + \sigma \lambda \tau_y .
\ee
The last term proportional to $\lambda$ anticommutes with the second and third terms, and thus gaps the Dirac points. Indeed, it can be recognized as the term which gaps a QBC~\cite{FradkinKivelson}. The nontrivial topology of the band structure immediately follows from the analysis in terms of a QBC point. A gapped QBC necessarily produces Chern bands~\cite{FradkinKivelson} with Chern number $C=\pm 1$. Since the spin up and down sectors have opposite signs of the gapping term ($\sigma \lambda \tau_y$), the ordering of the Chern bands is reversed from one spin sector to the other.  

\item {\it Topological gap closing transition at $X$ (Y).}

Next, consider an expansion around $\bQ_X=(\pi,0)$. Following the procedure used to derive Eq. \eqref{eq:X_exp_no_SOC}, we find:
\be
\mathcal H^{\up,\down}(\bQ_X+\bq) = -t_2(q^2_x-q^2_y )\mathbb{1} + [\pm J N +t_d(4-q^2) ]\tau_z + 2 t_1  q_x \tau_x  \pm  \frac12 \lambda q_y \tau_y.
\ee
As discussed after Eq. \eqref{eq:X_exp_no_SOC}, the Hamiltonian is diagonal along the line $q_x=0$ (corresponding to $MX$) in the absence of SOC ($\lambda=0$). Moreover, it is gapless for one of the spin species at $X$ when $|J|N = 4|t_d|$. In the case $J, t_d>0$, which we assume for now, the gapless point emerges only in the spin-$\down$ sector. At this gapless point, the Hamiltonian without SOC is given by \eqref{eq:X_exp_no_SOC_gapless}. When SOC is included, the Hamiltonian at the gapless point $J N= 4t_d$ (assuming still $J, t_d>0$) takes the form
\be
\mathcal H^{\down}(\bQ_X+\bq) = -t_2(q^2_x-q^2_y )\mathbb{1} -t_d q^2 \tau_z + t_1 2 q_x \tau_x  +  2 \lambda q_y\tau_y,
\ee
which is the Hamiltonian of a massless Dirac fermion. We conclude that as $J N$ increases from $J N < 4t_d$ to $J N> 4t_d$, the spin-$\up$ sector goes through a topological gap closing transition. At this transition, bands with opposite inversion $\mathcal{I}$ eigenvalues invert, as can be inferred from the discussion in Sec.~\ref{ssec:sym-2D}. This is a signal that the (parity of the) Chern number changes at the transition. Note that if the transition at $X$ occurs in the spin-$\down$ sector, then it occurs in the spin-$\up$ sector at $Y$. Hence, the Chern number changes in both spin sectors and it can be shown that the change is equal in magnitude but opposite in sign. 

\end{itemize}

\subsection{Quantum Spin Hall Effect \label{ssec:SOC-Nz-QSHE}}
\begin{figure}[h]
	\includegraphics[width=0.7\textwidth]{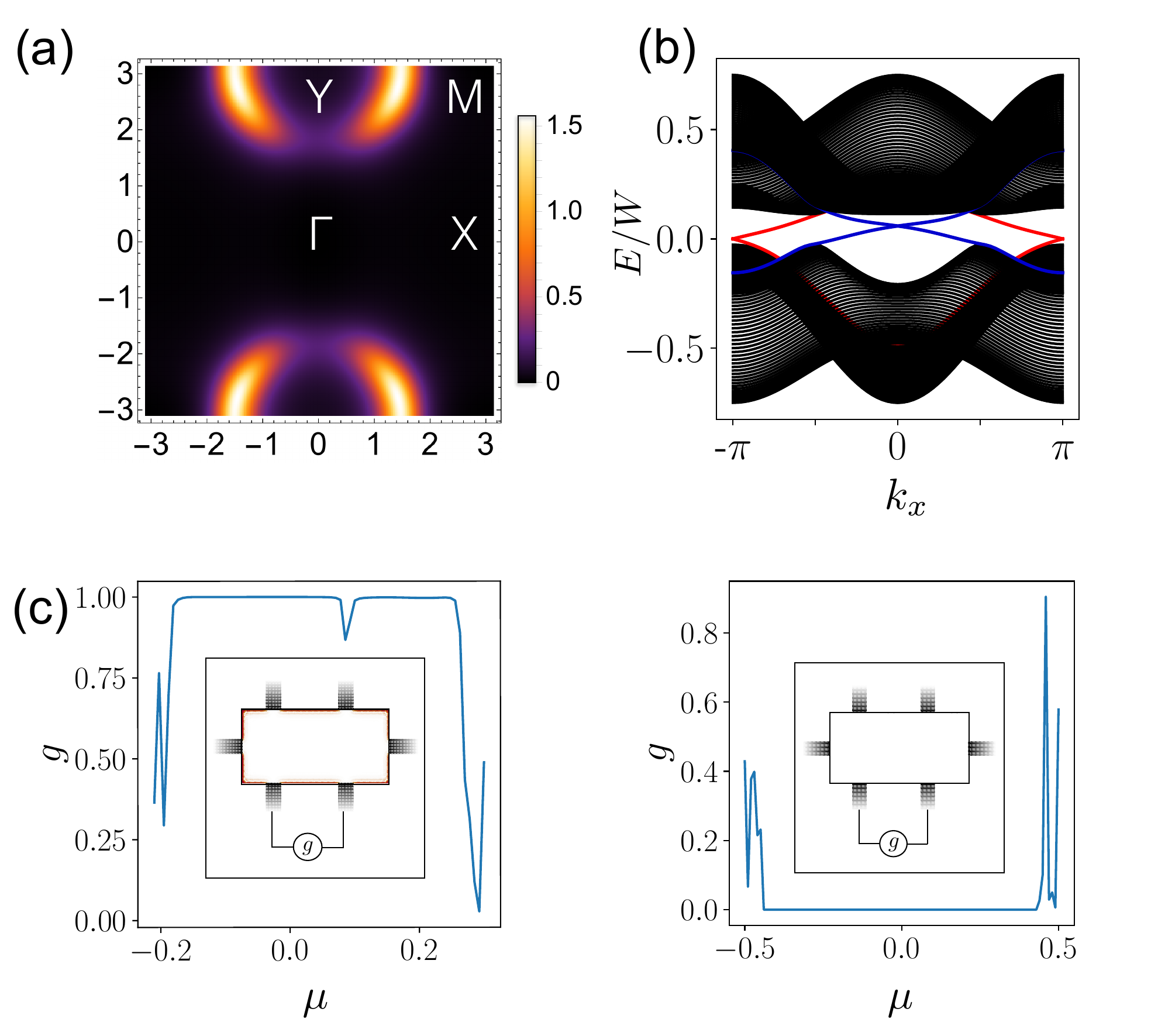}
	\caption{Results of the numerical simulation of the 2D altermagnet with SOC and out-of-plane N\'eel vector in the topological regime of mirror Chern insulator. The model parameters are the same as in Fig.~3(a) of the main text. (a) Spread of the Berry curvature (in color code) over the Brillouin zone for the lowest spin-up band. This band has mirror Chern number $1$. (b) Tight-binding calculation of the band dispersion in the stripe geometry (50 unit cell width) demonstrating edge modes shown in red and blue for spin-up and spin-down states, respectively. The energy is shown in units of $W = 8 t_d$. (c) Two-terminal conductance in the topological (left) and trivial (right) regimes as a function of chemical potential swept over the gap. The parameters in the left and right panels are the same as in Fig.~3(a) and (c), respectively. The geometry is shown in the inset, which also demonstrates the local density of states accumulated near the sample boundary in the topological phase. 
	}
	\label{fig:SM-Kwant}
\end{figure}

As discussed in the main text, the 2D model at small values of $N$ with \textit{out-of-plane} N\'eel vector develops a topological gap in the presence of spin-orbit coupling. The two lowest spin-up(down) bands are expected to have Chern numbers $\mathcal{C} = \pm 1$. To confirm this, we calculate numerically the Berry curvature in the Brillouin zone using a $200\times200$ mesh and employing the determinant trick \cite{Fukui-Berry, Vanderbilt-BerryBook}. Our calculations show that when $\lambda$ is increased, the Berry curvature, which is initially concentrated near the Dirac crossings along the $MX$ and $MY$ edges, is redistributed over the BZ, forming the peculiar circular pattern shown in Fig.~\ref{fig:SM-Kwant}(a). Note that the Chern number, which is the integrated Berry curvature, remains constant. 

Chern bands with opposite topological numbers for spin-up and spin-down bands are expected to lead to a version of the quantum spin Hall effect (QSHE) with counter-propagating spin-up/down edge states. To verify this prediction, we performed a numerical tight-binding simulation with the Kwant package \cite{Kwant-package}.  
The results are presented in Fig.~\ref{fig:SM-Kwant}(b,c). In panel (b), we show the band dispersions in the topological phase in the stripe geometry with 50 unit cells in width. Mid-gap topological edge modes are clearly visible, with one mode per edge and spin projection. Spin-up and spin-down boundary modes are highlighted in red and blue, while the bulk spectrum is shown in black regardless of the spin. 
In panel (c), we plot the two-terminal dimensionless electrical conductance $g$ between adjacent leads, revealing quantization $g = 1$ in the topological phase (left panel) and vanishing conductance otherwise (right panel). We used the same parameters as given in Figs. 3~(a) and (c) of the main text. The setup is shown in the insets, where we also plot the local density of states (LDOS) for a mid-gap chemical potential. In the topological regime, the edge modes demonstrably contribute to the LDOS near the edges of the slab. 

\subsection{Electronic structure for in-plane N\'eel vector \label{ssec:SOC-Nxy}}

We now consider the electronic properties of our model when the N\'eel vector lies in the plane. Based on the general analysis within the framework of the GL theory, as described in Sec.~\ref{sec:GL}, two cases need to be distinguished: (i) a N\'eel vector along the principal axes, and (ii) a N\'eel vector along the body diagonal, e.g. $\hat n = (1,1)/\sqrt{2}$. Here we consider these two cases separately.

\begin{itemize}

\item {\it N\'eel vector along the principal axes.} 

When the N\'eel vector points along one of the principal axes, say $\hat x$, the remaining symmetries are $\mathcal C_{2x}$, $\mathcal{M}_x$, and $\mathcal I$. While $\mathcal C_{2x}$ leaves the $MY$ line invariant, $\mathcal{M}_x$ leaves the $MX$ line invariant. The Brillouin zone boundaries therefore remain lines of high symmetry, which implies that the Dirac crossings are not gapped by the SOC. 

To see this, consider the line $MX$ parametrized by $(\pi, k_y)$. On this line, the representation of the mirror symmetry $\mathcal{M}_x$ is given by $U_{\mathcal{M}_x} = i \tau_z\sigma_x$, which must commute with the Hamiltonian. It is straightforward to check that it indeed does. It is therefore possible to diagonalize $U_{\mathcal{M}_x}$ and the Hamiltonian simultaneously, and to re-express the Hamiltonian in a basis of eigenstates of $U_{\mathcal{M}_x}$. This results in a block diagonal structure of the Hamiltonian with two $2\times 2$ blocks labeled by the mirror eigenvalues $\pm i$ of $U_{\mathcal{M}_x}$.  We denote the blocks $\mathcal H_\pm (k_y) $ and find that they are given by 
\be
\mathcal H_\pm (k_y) = \left( 4 t_2 \sin^2 \frac{k_y}{2} \pm J N \right) \mathbb{1} + \left( 4 t_d \cos^2 \frac{k_y}{2}\pm \lambda_{\text{FM}}N \right) \mu_z + \lambda \sin \frac{k_y}{2}\mu_y.
\ee
Here we have introduced a set of Pauli matrices $(\mu_x,\mu_y,\mu_z)$ which span the Hamiltonian in the mirror subspaces. That is to say, we introduce this additional set of Pauli matrices to write the $2\times 2$ blocks labeled by the mirror eigenvalues of $U_{\mathcal{M}_x}$. Note also that we have included the allowed magnetization term of Eq.~\eqref{H_AM-2D_SOC_inplane}.

To proceed, let us assume (without loss of generality) that $JN>0$. Point nodes on the $MX$ line then occur when the lower spectral branch of the $+i$ mirror sector crosses the upper branch of the $-i$ mirror sector. The precise condition for such a crossing is
\be
2JN  = \sqrt{\left( 4 t_d \cos^2 \frac{k_y}{2}+ \lambda_{\text{FM}}N \right)^2 +  \lambda^2 \sin^2 \frac{k_y}{2}} + \sqrt{\left( 4 t_d \cos^2 \frac{k_y}{2}- \lambda_{\text{FM}}N \right)^2 +  \lambda^2 \sin^2 \frac{k_y}{2}},
\ee
which defines a single constraint with one free parameter $k_y$. Such constraint is generically satisfied at a point. An argument of this kind, which examines the number of constraints and number of free parameters is referred to as a co-dimension argument.

To examine in more detail when the constraint is satisfied, let us set $\lambda_{\text{FM}}=0$ (i.e., ignore the overall magnetization), in which case the constraint simplifies and reads as
\be
J N = 4\sqrt{ t^2_d \cos^4 \frac{k_y}{2}+ \lambda^2 \sin^2 \frac{k_y}{2}}. \label{criterion}
\ee
It is instructive to compare this situation with the $\lambda=0$ case, i.e., the case without SOC. In this situation, the criterion (\ref{criterion}) is satisfied, and thus nodal crossings are present, provided that
\be
0 < \frac{J N}{4 |t_d|} < 1,
\ee 
Since the lower bound takes place for $k_y=\pi$ and the upper bound, for $k_y = 0$, the nodal crossings are created at $M$ and are removed at $X$, in full agreement with the previous analysis without SOC. When $\lambda$ is made nonzero the situation changes. The criterion (\ref{criterion}) is now satisfied provided that
\be
\mathrm{min}\left( \frac{\lambda}{|t_d|} \sqrt{1-\frac{\lambda^2}{4 t_d^2}}, \, 1 \right) < \frac{J N}{4 |t_d|} < \mathrm{max}\left( 1, \, \frac{\lambda}{|t_d|} \right)
\ee
Interestingly, the lower bound is now met either at $k_y= 2 \arccos (\frac{\lambda}{\sqrt{2} |t_d|})$, for $\lambda < \sqrt{2} |t_d|$ or $k_y=0$, for $\lambda > \sqrt{2} |t_d|$. Conversely, the upper bound is met either at $k_y= 0$, for $\lambda < |t_d|$, or $k_y=\pi$, for $\lambda > |t_d|$. Therefore, the number of crossings depends on $\lambda$. There are two different regimes: for $\lambda < \sqrt{2} |t_d|$, there is only one crossing if
\be
\frac{\lambda}{|t_d|} < \frac{J N}{4 |t_d|} < 1,
\ee
and two crossings if
\be
\frac{\lambda}{|t_d|} \sqrt{1-\frac{\lambda^2}{4 t_d^2}} < \frac{J N}{4 |t_d|} < \mathrm{min}\left( 1, \, \frac{\lambda}{|t_d|} \right).
\ee
On the other hand, for $\lambda > \sqrt{2} |t_d|$, there is only one crossing provided that
\be
1 < \frac{J N}{4 |t_d|} < \frac{\lambda}{|t_d|}.
\ee
Note that this second crossing survives even in the regime where there is no crossing in the absence of SOC.

\item {\it N\'eel vector along $\hat n = (1,1)/\sqrt{2}$.}

Finally, we consider the case of a N\'eel vector along the body diagonal, i.e., along $\hat n = (1,1)/\sqrt{2}$. In this case the lines $MX$ and $MY$ are not invariant under the remaining physical symmetries, which leads to an immediate gapping of the Dirac crossings. 
 
To see this in more detail, we set $k_x=\pi$ and find that the full Hamiltonian on the $k_x=\pi$ lines is given by
\be
\mathcal H(\pi,k_y) = 4 t_2 \sin^2 \frac{k_y}{2} \tau_0 + 2  t_d \cos^2 \frac{k_y}{2} \tau_z + \lambda \sin \frac{k_y}{2} \tau_y \sigma_z+\frac{JN}{\sqrt{2}} \tau_z(\sigma_x+\sigma_y) +\frac{\lambda_{\text{FM}} N}{\sqrt{2}} (\sigma_x-\sigma_y)  .
\ee
Importantly, here we have included the allowed magnetization term of Eq.~\eqref{H_AM-2D_SOC_inplane} with coupling $\lambda_{\text{FM}} $. The importance of this term can be seen as follows. If we were to ignore it, we would be left with a Hamiltonian that commutes with $\tau_z(\sigma_x+\sigma_y)/\sqrt{2}$ and therefore be block diagonalized by expressing it in terms of the eigenvectors of $\tau_z(\sigma_x+\sigma_y)/\sqrt{2}$. We would then face a situation similar to the case of a N\'eel vector along $\hat x$, where a crossings of bands from different blocks is stable. The coupling $\lambda_{\text{FM}} N (\sigma_x-\sigma_y) /\sqrt{2}$ does not commute with $\tau_z(\sigma_x+\sigma_y)/\sqrt{2}$, however, and therefore does not block diagonalize and indeed couples the blocks. This invalidates the co-dimension argument and leads to a gapping of the Dirac points. We thus find that once the allowed canting of moments is taken into account, the Dirac points gap.

\end{itemize}

\section{Altermagnetism on the 3D rutile lattice without SOC \label{sec:3D-model}}

In the main text we introduced a generic tight-binding model for altermagnetism in a class of systems defined by the rutile lattice structure, such as RuO$_2$ and MnF$_2$. In this Supplemental Section we provide details of the construction and analysis of the 3D rutile model, which is based on Ref. \cite{McClarty_LongSpinGroups}.

We proceed in the same manner as in the case of the 2D model and write the Hamiltonian in the form
\be
H = \sum_\bk c^\dagger_\bk \mathcal H(\bk) c_\bk, \qquad c^\dagger_{\bk} = \begin{pmatrix} c^\dagger_{ \bk A \up}  & c^\dagger_{\bk B \up}&  c^\dagger_{ \bk A \down}  & c^\dagger_{\bk B \down}\end{pmatrix}. \label{ham-3D}
\ee
In this section we consider the case when SOC is absent, such that the Hamiltonian $ \mathcal H(\bk)$ consists of the two terms
\be
\mathcal H = \mathcal H_t + \mathcal H_{\text{AM}} \equiv \mathcal H_0.  \label{H_0-3D}
\ee
The sublattices $A$ and $B$ correspond to the two magnetic Ru or Mn atoms in the unit cell, sitting at Wyckoff positions $(0,0,0)$ and $(\frac12,\frac12,\frac12)$.  We recall that, in contrast to the 2D model, the origin of the coordinate system is placed at one of the magnetic atoms.

\subsection{Construction of the Hamiltonian \label{sec:RuO}}

To obtain a generic model for the rutile class, we follow the approach of Refs.~\onlinecite{McClarty_LongSpinGroups,McClarty-AltermagnetismGL} and include hopping terms up to third-nearest neighbors. As pointed out in Ref.~\onlinecite{McClarty_LongSpinGroups}, including third-nearest neighbor hopping is necessary to ensure that the model reflects the true symmetry of the crystal. In particular, hopping anisotropies arising from the distinct crystalline environment of the $A$ and $B$ sublattice only show up at third-nearest neighbors. This can also be inferred from the fact that the point-group-invariant combination $\tau_z d_{xy}$ (see Table~\ref{tbl:irreps}) can be implemented only with the help of third-nearest neighbor hopping.
 
The first-, second- and third-nearest neighbor hoppings are denoted $t_1$, $t_2$ ($t'_2$), and $t_{3d}$, respectively. Note that second-nearest neighbor hopping along the $ab$ plane ($t_2$) is generally different from hopping along the $c$ axis ($t'_2$). Furthermore, we only include the anisotropic third-nearest neighbor hopping $t_{d}$, which has a $d_{xy}$ symmetry, since the isotropic term does not add new features to the system. The Hamiltonian $\mathcal H_t (\bk)$ then takes the form
\be
\mathcal H_t (\bk) = -8 t_1  \cos \frac{k_x}{2}  \cos \frac{k_y}{2}\cos \frac{k_z}{2} \tau_x -2 [ t_2(\cos k_x + \cos k_y)  + t'_2 \cos k_z] \tau_0 - 4  t_{d}  \sin k_x  \sin  k_y \tau_z.
\ee
Using the abbreviated notation introduced in the main text, 
\be
c_{i/2} \equiv \cos \frac{k_i}{2} , \qquad c_i \equiv \cos k_i , \label{def-cos-sin}
\ee
it takes the form
\be
\mathcal H_t (\bk) = -8 t_1 c_{x/2}c_{y/2}c_{z/2} \tau_x -2 [ t_2(c_x+c_y)  + t'_2 c_z] \tau_0 - 4 t_{d}  s_xs_y \tau_z.
\ee
For simplicity, in what follows we will set $t_2=t'_2$, which does not affect our results and conclusions. As it was the case for the 2D model, it will be useful to define the form-factor functions
\be
s_\bk =   8 \cos \frac{k_x}{2}  \cos \frac{k_y}{2}\cos \frac{k_z}{2} , \qquad s^{\prime}_\bk= 2(\cos k_x + \cos k_y+\cos k_z), \qquad d_\bk =  4 \sin k_x  \sin  k_y , \label{form-factors-3D}
\ee

Next we include the altermagnetic Hamiltonian. The form of $ \mathcal H_{\text{AM}}(\bk) $ is very similar to \eqref{H_AM-2D} and only differs in the magnetic octupole term:
\be
 \mathcal H_{\text{AM}}(\bk)  = J\tau_z  \bN \cdot  \bsigma + J' \sin k_x  \sin  k_y  \bN \cdot  \bsigma .\label{H_AM-3D}
\ee
As in the 2D case, it is not essential to retain the symmetry-allowed $J'$ term, and we therefore keep only the $J$ term in the subsequent analysis (see also the discussion in Sec.~\ref{ssec:ham}).

\subsection{Symmetries of the Hamiltonian \label{ssec:sym-3D}}

Next, we present an analysis of the relevant symmetry-mandated degeneracies in the absence of SOC. This discussion follows the approach of Sec.~\ref{ssec:sym-2D} and makes use of the same notation. Here we focus only on the Brillouin zone (BZ) boundaries, where degeneracies of interest arise. 

\begin{itemize}

\item {\it The BZ boundary plane $k_z=\pi$.} 

We begin with the $k_z=\pi$ plane given by $\bQ_{k_z=\pi}=(k_x,k_y,\pi)$. This BZ boundary is invariant under the mirror symmetry $ \{ \mathcal M_{z} |000 \} $, which implies that bands can be labeled by mirror eigenvalues. Note that 
\be
\mathcal M_z \bQ_{k_z=\pi} = \bQ_{k_z=\pi}- \bfg_3,
\ee
where $ \bfg_3 = 2\pi/c$ is a reciprocal lattice vector. Therefore, the symmetry representation in the space of Bloch states takes the form
\be
U_{\mathcal M_z} = V_{-\bfg_1} = \tau_z. \label{sym-M_z-3D}
\ee
Similar to what was found in the 2D model on the high symmetry lines, this implies that bands from different sublattices have opposite mirror eigenvalues and therefore cannot mix on the $k_z=\pi$ plane. The Hamiltonian must therefore be diagonal on the $k_z=\pi$ plane.

\item {\it The BZ boundary planes $k_x=\pi$ and $k_y=\pi$.} 

Next, we consider the $k_x=\pi$ and $k_y=\pi$ planes. Since they are related by symmetry, it is sufficient to consider the $k_x=\pi$ plane. Our goal is to show that bands must be twofold degenerate on the entire $k_x=\pi$ plane when the symmetry $\mathcal T \{ \mathcal C_{2x} |  \tfrac12 \tfrac12 \tfrac12  \}$ is present, i.e., the product of time-reversal ($\mathcal T$) and the non-symmorphic $ \{ \mathcal C_{2x} |  \tfrac12 \tfrac12 \tfrac12  \}$, which corresponds to a two-fold rotation followed by a half translation. Note that this product of symmetries indeed leaves the BZ boundary defined by $\bQ_{k_x=\pi}=(\pi,k_y,k_z)$ invariant. To prove that the presence of $\mathcal T \{ \mathcal C_{2x} |  \tfrac12 \tfrac12 \tfrac12  \}$ implies a twofold degeneracy of the energy bands, consider first the algebraic space group relation
\be
 \{ \mathcal C_{2x} |  \tfrac12 \tfrac12 \tfrac12  \}  ^2 = \{ \mathcal C^2_{2x} |  100  \} = \mathcal R  \{ \mathbb{1} |  100  \},
 \ee
 where $\mathcal R$ denotes a rotation by $2\pi$. In the absence of SOC (i.e., when the spins rotate independently of the lattice), one simply has $\mathcal R = 1$. Similarly, in the absence of SOC $\mathcal T$ is just complex conjugation and therefore $\mathcal T^2=1$. It follows that
 \be
 ( \mathcal T \{ \mathcal C_{2x} |  \tfrac12 \tfrac12 \tfrac12  \} ) ^2 =  \{ \mathbb{1} |  100  \}
 \ee
 The representation of any pure translation $\{ \mathbb{1} |  {\bf t}  \}$ is $e^{-i \bk \cdot {\bf t}}$, which means that the Block representation of $ \{ \mathbb{1} |  100  \}$ equals $e^{-i k_x}$. On the $k_x=\pi$ plane this reduces to $-1$, which in turn implies that $ \mathcal T \{ \mathcal C_{2x} |  \tfrac12 \tfrac12 \tfrac12  \} $ is an anti-unitary symmetry (due to $\mathcal T$) which squares to $-1$. Kramers theorem then applies, since the $k_x=\pi$ plane is invariant under $\{ \mathcal C_{2x} |  \tfrac12 \tfrac12 \tfrac12  \} $, thus mandating a manifest twofold degeneracy on the $k_x=\pi$ plane.
 
For our problem this result has the following implications: in the absence of magnetic order, all bands are fourfold degenerate due to the trivial additional spin degeneracy. In the presence of a collinear (alter)magnetic order, the twofold rotation $ \{ \mathcal C_{2x} |  \tfrac12 \tfrac12 \tfrac12  \}$ symmetry is broken, which allows for a splitting of the bands. Since the splitting is opposite for the two spin species, a fourfold-degenerate state reduces to two twofold-degenerate states.

\end{itemize}

\subsection{Analysis of the Hamiltonian and the spectrum \label{ssec:ham-analysis-3D}}

Without SOC, we are free to set the N\'eel vector along the $\hat z$ direction (without loss of generality). The Hamiltonian takes the form
\be
\mathcal H^{\up,\down}_0(\bk) = \begin{pmatrix} \pm JN -t_2 s'_\bk - t_{d} d_\bk  & -t_1 s_\bk  \\ -t_1 s_\bk  & \mp JN -t_2 s'_\bk + t_{d} d_\bk  \end{pmatrix} =  -t_2 s'_\bk \tau_0 -t_1 s_\bk \tau_x -(t_{d} d_\bk \pm JN)\tau_z ,
\ee
with the form-factor functions $s_\bk$, $s'_\bk$, and $ d_\bk$ given by \eqref{form-factors-3D}. Note that the structure of the Hamiltonian is identical to that of the 2D model; the only meaningful difference lies in the $d$-wave form factor $d_\bk$, which has $k^2_x-k^2_y$ symmetry in the 2D case and $k_xk_y$ symmetry in the 3D case. 

The energy spectrum is then given by:
\be
\mathcal E^{\sigma}_\pm(\bk) =  -t_2 s'_\bk  \pm \sqrt{t^2_1 s^2_\bk + (t_{d} d_\bk -\sigma JN)^2}.
\ee
As in the 2D model, this form highlights that a spin splitting can only occur when $t_{d}$ is nonzero, and that it vanishes whenever $d_\bk $ has nodes. In this 3D model, the nodes of $d_\bk \sim \sin k_x \sin k_y $ occur on planes in momentum space.

We now analyze the Hamiltonian and its spectrum on special planes of high symmetry, specifically the $k_z=\pi$ and $k_x=\pi$ planes. 

\begin{itemize}

\item {\it Weyl line nodes on the $k_z=\pi$ plane.}

We begin with the $k_z=\pi$ plane. As described in Sec.~\ref{ssec:sym-3D}, the representation of $\mathcal M_z$ on the $k_z=\pi$ plane is given by $U_{\mathcal M_z} = \tau_z$, see Eq.~\eqref{sym-M_z-3D}, which implies that the Hamiltonian must commute with $\tau_z$ and thus must be diagonal. When $\bk$ is restricted to the $k_z=\pi$ plane, the Hamiltonian becomes
\be
\mathcal H^{\sigma}_0(\bk) = \begin{pmatrix}  \sigma JN -t_2 s'_\bk - t_{d} d_\bk  & 0  \\ 0  &  -\sigma JN -t_2 s'_\bk + t_{d} d_\bk  \end{pmatrix} .
\ee
It follows from this form of $\mathcal H^{\sigma}_0(\bk) $ that bands of the same spin must cross on lines in momentum space. To see this, consider for simplicity $JN>0$; the case $JN<0$ is essentially equivalent. Furthermore, consider the $\sigma = \up$ sector. The bands in this sector are degenerate when $JN - t_{d} d_\bk = 0$, which is satisfied in the quadrants where $t_d d_\bk$ is positive as long as $ JN < 4|t_d|$. Since this condition defines one constraint and we have two parameters $(k_x,k_y)$, the degeneracy occurs on a line. We emphasize that the bands that cross have opposite sublattice index and therefore opposite mirror eigenvalue. 

Now consider the $\sigma = \down$ sector. The condition for degeneracy is $-JN - t_{d} d_\bk = 0$, which is satisfied in the quadrants where $t_d d_\bk$ is negative (again as long as $ JN < 4|t_d|$). This demonstrates the existence of spin-polarized line nodes in the $k_z=\pi$ plane. The line nodes of opposite spins are related by fourfold rotations, as expected for a $d$-wave altermagnet. They are shown by the red and blue loops in Fig. 4(b) of the main text.

\item {\it The $k_x=\pi$ plane: a Weyl nodal plane.}

Now consider the $k_x=\pi$ plane. In Sec.~\ref{ssec:sym-3D} we presented a general argument for a symmetry-mandated fourfold degeneracy on the entire plane without magnetism. This is clearly reflected in the form of the Hamiltonian. In particular, excluding magnetism for the moment, we readily find that on the $k_x=\pi$ plane $\mathcal H_t$ reduces to
\be
\mathcal H_t(\pi,k_y,k_z) = - 2 t_2 (c_y+c_z) \tau_0 ,
\ee
and is thus proportional to the identity. Expanding in small momentum $q_x$ away from the $k_x=\pi$ plane, we find that the Hamiltonian takes the form
\be
\mathcal H_t(\pi+q_x,k_y,k_z) = - 2 t_2 (c_y+c_z) \tau_0 +  4q_x(2 t_1c_{y/2}c_{z/2}\tau_x -t_d s_y\tau_z ),
\ee
which shows that the energy dispersion is linear in $q_x$. The $k_x=\pi$ plane may therefore be called a Dirac nodal plane, as was first recognized in Ref.~\onlinecite{Sun:2017p235104}. Note that the velocity characterizing the linear dispersion is dependent on $(k_y,k_z)$, and so are the conduction and valence band eigenstates. 

In the presence of magnetism the Hamiltonian $\mathcal H_0 = \mathcal H_t + \mathcal H_{\text{AM}} $ on the $k_x=\pi$ plane becomes
\be
\mathcal H^{\sigma}_0(\pi,k_y,k_z) = \begin{pmatrix} \sigma JN  - 2 t_2 (c_y+c_z)   & 0  \\ 0  & -\sigma JN  - 2 t_2 (c_y+c_z)   \end{pmatrix} .
\ee
The fourold degeneracy is now split into twofold degeneracies, which implies that the Dirac nodal plane turn into a Weyl nodal plane. 

Finally, we plot the dispersion of the 3D model without spin-orbit coupling for several values of the on-site Kondo term $J N$ in Fig.~\ref{fig:3D-extra}.
\begin{figure}[h]
\includegraphics[width=\textwidth]{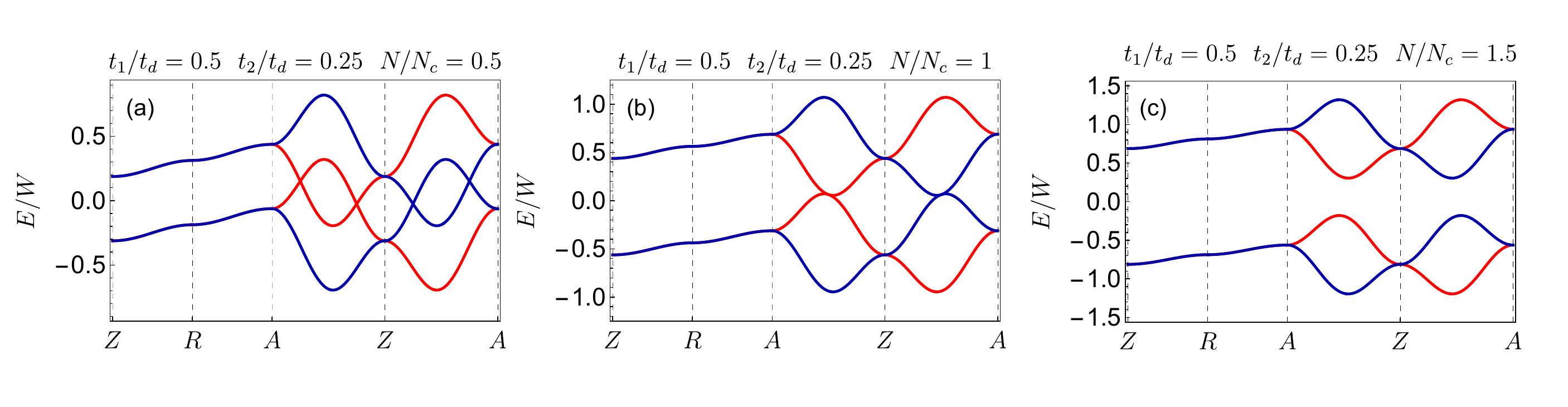}
	\caption{Electron dispersion in the 3D model for different values of the on-site Kondo term $J N$. Panel (a) corresponds to Fig.~1(a) of the main text. As can be seen in the panels (a)-(c), increase of the $JN$ parameter leads to the disappearance of the Weyl nodal lines at $N = N_c = |4 t_d / J|$}
	\label{fig:3D-extra}
\end{figure}
This figure demonstrates that  an increase of the Kondo coupling leads to the disappearance of the Weyl nodal lines at $N = N_c = |4 t_d / J|$. Thus, the Weyl lines are symmetry protected for low and moderate values of $J N$ but not symmetry-enforced, and do not exist for larger values of $J N$. Importantly, the $J N$ parameter can be decreased by heating the system to bring it closer to the phase transition, where the staggered magnetization $N$ is small. 
This behavior of the Weyl nodal lines is similar to the behavior of the Dirac crossings in the 2D model. However, contrary to the Dirac points, the Weyl nodal lines are stable with respect to the inclusion of spin-orbit coupling. 

\end{itemize}

\section{Altermagnetism on the 3D rutile lattice with SOC \label{sec:3D-model-SOC}}

We now turn to the 3D model for the rutile lattice in the presence of SOC. As in the 2D case, the Hamiltonian of Eq.~\eqref{ham-3D} acquires an additional spin-orbit term $ \mathcal H_{\text{SO}} $, such that the full Hamiltonian becomes
\be
\mathcal H = \mathcal H_t + \mathcal H_{\text{AM}} + \mathcal H_{\text{SO}}=  \mathcal H_{0} + \mathcal H_{\text{SO}}.  \label{H-3D-SOC}
\ee
Furthermore, the form of $\mathcal H_{\text{AM}} $ is altered due to the locking of the N\'eel vector to the lattice promoted by SOC. 

Consider first the form of $ \mathcal H_{\text{SO}}$. Using representation theory (see Table~\ref{tbl:irreps}), we find two symmetry-allowed spin-orbit terms, which, using the abbreviations defined in \eqref{def-cos-sin}, are given by
\be
 \mathcal H_{\text{SO}}(\bk) =\lambda_1 c_{x/2}c_{y/2}c_{z/2} (c_x-c_y)\tau_y \sigma_z + \lambda_2 s_{z/2} (s_{x/2}c_{y/2}\tau_y \sigma_x - s_{y/2}c_{x/2}\tau_y \sigma_y). \label{H_SO-3D}
\ee
We emphasize that these terms are allowed by the symmetries of the space group $P4_2/mnm$ and do not in any way require magnetic order. 

We briefly comment on the nature of these two terms. The first term can be understood as follows. Consider the product of Pauli matrices $\tau_y \sigma_z$. It is time-reversal and inversion invariant and transforms as the $B_{1g} \times A_{2g} = B_{2g}$ irreducible representation. To obtain a fully symmetric SOC term it would have to be multiplied with a $\bk$-dependent form factor with $B_{2g}$ symmetry which is even in $\bk$ and connects different sublattices due to the presence of $\tau_y$. The first tight-binding implementation of this term appears in the fourth-nearest neighbor spin-dependent hopping with form-factor given by $c_{x/2}c_{y/2}c_{z/2} (c_x-c_y)$. Due to the factor $c_{x/2}c_{y/2}c_{z/2}$, this term vanishes on the entire BZ boundary.  

To understand the presence of the second term, consider the following group theoretical argument. Since Pauli matrices $(\sigma_x,\sigma_y)$ and the form factors $(s_{z/2} s_{x/2}c_{y/2}, s_{z/2} s_{y/2}c_{x/2})$ have $E_g$ symmetry, products of these will generate the following irreps, $E_g \otimes E_g = A_{1g} \oplus A_{2g} \oplus B_{1g} \oplus B_{2g}$. Since the form factors are even in momentum, such products are necessarily odd under time-reversal. Thus, to appear in the Hamiltonian, this term must be combined with terms that are off-diagonal in sublattice space and described by $\tau_y$. Since $\tau_y$ has $ B_{1g}$ symmetry, we must look for the corresponding term in the product decomposition, which gives the last term in \eqref{H_SO-3D}. 

This group-theoretical argument does not, however, reveal where the form factors come from. To understand this, consider the rotated Pauli matrices defined as
\be
\sigma' _x = \hat n_1 \cdot \bsigma, \qquad \sigma' _y = \hat n_2 \cdot \bsigma, \qquad \text{with} \qquad \hat n_1 = \frac{1}{\sqrt{2}} (1,1,0), \qquad \hat n_2 = \frac{1}{\sqrt{2}} (-1,1,0),
\ee
which correspond to the in-plane coordinate axes rotated by $\pi/4$. In such a rotated frame, the appropriate $d$-wave form factors are
\be
d'_{xz} = 4 \sin  \frac{k_z}{2}  \sin \frac{k_x+k_y}{2}, \qquad d'_{yz} = 4 \sin  \frac{k_z}{2}  \sin \frac{k_y-k_x}{2}. 
\ee
These form factors clearly correspond to nearest neighbor hopping between the sublattices. Consider now the mirror symmetry $\{ M_{-x+y} | 000  \}$. Under this mirror operation one has $\sigma' _x \rightarrow - \sigma' _x$, $\sigma' _y \rightarrow  \sigma' _y$, $d'_{xz}  \rightarrow  d'_{xz} $, and $d'_{yz}  \rightarrow  -d'_{yz} $. Since $\tau_y$ is even under this mirror (the sublattices are not exchanged), the correct symmetric combination of form factors and spin Pauli matrices is $d'_{xz} \sigma' _y +d'_{yz}\sigma' _x$. Using the definitions of the rotated matrices and form factors, it is then straightforward to show that
\be
d'_{xz} \sigma' _y +d'_{yz}\sigma' _x \sim  s_{z/2} (s_{x/2}c_{y/2}\tau_y \sigma_x - s_{y/2}c_{x/2}\tau_y \sigma_y).
\ee

Having discussed the allowed SOC terms, let us now briefly examine the form of $\mathcal H_{\text{AM}} $ in the presence of SOC. This discussion is fully analogous to the discussion in Sec.~\ref{sec:2Dmodel-SOC} in the context of the 2D model. Again, per the analysis of Sec.~\ref{sec:GL}, we distinguish $N_z$ and $(N_x,N_y)$, i.e., an out-of-plane N\'eel vector from an in-plane N\'eel vector. In the case of the former, $\mathcal H_{\text{AM}} $ takes the form
\be
\mathcal H_{\text{AM}} (\bk) =  JN_z \tau_z \sigma_z + J' N_z \sin k_x\sin k_y \sigma_z, \label{H_AM_3D_SOC}
\ee
whereas in the case of an in-plane N\'eel vector, $\mathcal H_{\text{AM}} $ takes the form
\be
\mathcal H_{\text{AM}} (\bk) =  J \tau_z (N_x \sigma_x + N_y\sigma_y )  + J' \sin k_x\sin k_y  (N_x \sigma_x + N_y\sigma_y ) + h (M_x \sigma_x + M_y\sigma_y).
\ee
As before, for an in-plane N\'eel vector the Hamiltonian acquires an additional term proportional to the magnetization, since the magnetization couples linearly to $(N_x,N_y)$. The direction of $\bM$ is determined by $\bN$ via Eq.~\eqref{F_NM-3D}, and therefore we may write the last term as $\lambda_{\text{FM}} (N_x \sigma_y + N_y\sigma_x)$. Recall that we always set $J'=0$.

\subsection{Symmetries in the presence of SOC \label{sec:sym-3D-SOC} }

\begin{figure}
	\includegraphics[width=\textwidth]{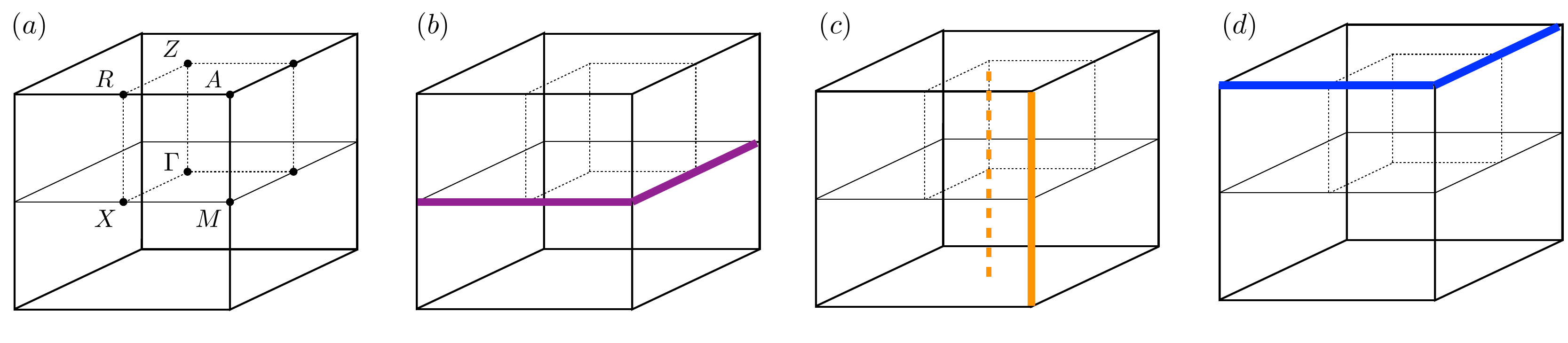}
	\caption{Points and lines of high symmetry of the 3D tetragonal BZ. (a) Points of high symmetry. (b) Lines of high symmetry in the $k_z=0$ plane. (c) Vertical lines of high symmetry including fourfold rotations. (d) Lines of high symmetry in the $k_z=\pi$ plane.}
	\label{fig:sym-3D}
\end{figure}

Before we turn to the analysis of the Hamiltonian in the presence of magnetic order in the next section, here we present a general analysis of the symmetries and their implications for the energy bands in the presence of SOC but without magnetic order. Our goal is to determine symmetry-enforced degeneracies that rely only on the algebraic structure of the space group in the absence of magnetism.

\begin{itemize}

\item {\it The BZ boundary plane $k_z=\pi$.} 

First consider the boundary plane $k_z=\pi$, which is left invariant by the mirror reflection $ \{ \mathcal M_z | 000 \}$. This mirror satisfied the algebraic space group constraint
\be
 \{ \mathcal M_z | 000 \}^2 = \mathcal R , \label{mirror_z-squared}
\ee
where $ \mathcal R$ is the rotation by $2\pi$, which equals $ \mathcal R=-1$ for spin-$1/2$ fermions. The representation of mirror $ \{ \mathcal M_z | 000 \}$ in the space of Bloch states at $k_z=\pi$ is equal to
\be
U_{\mathcal M_z} = V_{-\bfg_3} \otimes (-i \sigma_z) = -i \tau_z \sigma_z, \label{U-mirror_z}
\ee
and the Hamiltonian $\mathcal H(\bk)$ must commute with $U_{\mathcal M_z} $ for $k_z=\pi$. Note that $U_{\mathcal M_z}$ indeed satisfies $U^2_{\mathcal M_z} =-1$. Since the Hamiltonian commutes with $U_{\mathcal M_z} $ it block diagonalizes into two blocks labeled by the mirror eigenvalues $\pm i$. Crossings between bands from different mirror sectors are then locally stable. 

\item {\it The BZ boundary lines on the $k_z=0$ plane.} 

Next, consider the BZ boundary lines on the $k_z=0$ plane, i.e., the lines for which either $k_x=\pi$ or $k_y=\pi$, while $k_z=0$. These are indicated by bold purple lines in Fig.~\ref{fig:sym-3D}(b). Here we show that in the absence of magnetism i.e., when time-reversal ($\mathcal T$) is present, all bands must be fourfold degenerate. This was first proven in Ref.~\onlinecite{Sun:2017p235104}; here we reproduce that argument and emphasize that manifest twofold degeneracies may remain when $\mathcal T$ is broken. 

Consider the mirror reflection $ \{ \mathcal M_{z} | 000 \} $ and let $\mathcal O$ be either $\mathcal M_x$ or $\mathcal C_{2y}$. The symmetries $ \{ \mathcal M_{z} | 000 \} $ and $\{ \mathcal O | \tfrac12 \tfrac12 \tfrac12 \}$ leave the $k_x=\pi$ line invariant. Furthermore, these symmetries have the algebraic space group relation
\be
\{ \mathcal O | \tfrac12 \tfrac12 \tfrac12 \}  \{ \mathcal M_{z} | 000 \} = \{ \mathcal O \mathcal M_z | \tfrac12 \tfrac12 \tfrac12 \} =\mathcal R  \{ \mathcal M_z \mathcal O | \tfrac12 \tfrac12 \tfrac12 \} = \mathcal R  \{ \mathbb{1} | 001 \}  \{ \mathcal M_{z} | 000 \}  \{ \mathcal O | \tfrac12 \tfrac12 \tfrac12 \} \label{X-M_z}
\ee
which implies that representations of these symmetries on the $k_z=0$ plane anti-commute. Let the representations be given by $U_{\mathcal M_z}$ and $U_{\mathcal O}$ and note that $U_{\mathcal M_z}$ must have eigenvalues $\pm i$ due to Eq. \eqref{mirror_z-squared}. If $\ket{\psi}$ is an eigenstate of the Hamiltonian and of $U_{\mathcal M_z}$ with eigenvalue $\pm i$, then $U_{\mathcal O}\ket{\psi}$ is an eigenstate of $U_{\mathcal M_z}$ with eigenvalue $\mp i$, and thus orthogonal to $\ket{\psi}$. Now consider the product of time-reversal and inversion, given by $ \mathcal T\{ \mathcal I | 000 \}$, which leaves the $(k_x,k_z) = (\pi,0)$ line invariant and commutes with $\{ \mathcal M_{z} | 000 \}$, 
\be
\big[\{ \mathcal M_{z} | 000 \}, \mathcal T\{ \mathcal I | 000 \}\big] =0 . \label{M_z-TI-commute}
\ee
Since $\mathcal T$ is anti-unitary, $\mathcal T U_{\mathcal I} \ket{\psi}$ is also an eigenstate of $U_{\mathcal M_z}$ with eigenvalue $\mp i$, and $\mathcal T U_{\mathcal I} U_{\mathcal O} \ket{\psi}$ is an eigenstate of $U_{M_z}$ with eigenvalue $\pm i$. Hence, $\ket{\psi}$ and $\mathcal T U_{\mathcal  I} U_{\mathcal O} \ket{\psi}$ are both eigenstates of $U_{\mathcal M_z}$ with eigenvalue $\pm i$. They are orthogonal if $ \mathcal T\{ \mathcal  I | 000 \}\{ \mathcal O | \tfrac12 \tfrac12 \tfrac12 \} $ is an anti-unitary symmetry which squares to $-1$ on the $(k_x,k_z) = (\pi,0)$ line. To show this, consider $\mathcal O=\mathcal M_x$ and evaluate
\be
(\mathcal T\{ \mathcal I | 000 \}\{ \mathcal M_x | \tfrac12 \tfrac12 \tfrac12 \} )^2=(\mathcal T\{ \mathcal C_{2x} | \tfrac{\bar 1}{2} \tfrac{\bar 1}{2}  \tfrac{\bar 1}{2}  \} )^2 =\mathcal T^2 \mathcal R \{ \mathbb{1} | \bar 100 \} \label{T.I.M_x_squared}
\ee
Since $\mathcal T^2=-1$ and $ \mathcal R=-1$ for spin-$1/2$ fermions, and $ \{ \mathbb{1} | \bar 100 \} $ has representation $e^{i k_x}$, we see that indeed $(\mathcal T U_{\mathcal I} U_{\mathcal O})^2=-1$.

We emphasize that the fourfold degeneracy on the $(k_x,k_z) = (\pi,0)$ line, as well as on the $(k_y,k_z) = (\pi,0)$ line, follows from the algebraic space group relations, and is independent of the specific model under considerations. In any model all bands must be fourfold degenerate. 

Clearly this argument is no longer valid when one of the symmetries is broken. For instance, when $\mathcal T$ is broken due to magnetism (while inversion is preserved) the fourfold degeneracy is generically lifted. Spatial symmetries, when present, can nonetheless mandate twofold degeneracies, as is clear from the analysis of Eq.  \eqref{X-M_z}. In addition, the mirror reflection $\{ \mathcal  M_{x} | \tfrac12 \tfrac12 \tfrac12 \} $ and twofold rotation$  \{ \mathcal C_{2y} | \tfrac12 \tfrac12 \tfrac12 \} $, which leave $(k_x,k_z) = (\pi,0)$ invariant, have the algebraic space group relation
\be
\{ \mathcal M_{x} | \tfrac12 \tfrac12 \tfrac12 \}  \{ \mathcal C_{2y} | \tfrac12 \tfrac12 \tfrac12 \} = \{ \mathcal M_{x}\mathcal C_{2y} |011   \} =\mathcal R  \{  \mathcal C_{2y}\mathcal M_{x} |011  \} = \mathcal R  \{ \mathbb{1} | 001 \}  \{ \mathcal C_{2y} |  \tfrac12 \tfrac12 \tfrac12  \}  \{ \mathcal M_{x} | \tfrac12 \tfrac12 \tfrac12 \} , \label{M_x-C_2y}
\ee
which similarly implies that their representations must anticommute on the $(k_x,k_z) = (\pi,0)$ line.

\item {\it The BZ boundary lines in the $k_z=\pi$ plane.} 

Instead of $k_z=0$, consider now the BZ boundary lines of the $k_z=\pi$ plane, shown by bold blue lines in Fig.~\ref{fig:sym-3D}(d). The crucial difference is that now the space group relations \eqref{X-M_z} and \eqref{M_x-C_2y} imply a \emph{commutation} relation rather than an anticommutation relation of the symmetries involved. The argument given previously therefore does not hold anymore. The only manifest (twofold) degeneracy can come from the product $ \mathcal T\{ \mathcal I | 000 \}$ which is anti-unitary and squares to $-1$. When time-reversal symmetry is broken due to magnetism, no degeneracies are generically expected. 

\item {\it The vertical fourfold rotation lines $\Gamma Z$ and $MR$.} 

Finally, consider the fourfold rotation lines $\Gamma Z$ and $MR$, which are indicated by orange dashed and bold lines, respectively, in Fig.~\ref{fig:sym-3D}(c). These lines are left invariant by $ \{ \mathcal C_{2z} | 000 \}$ as well as $\{ \mathcal O | \tfrac12 \tfrac12 \tfrac12 \}$, where $\mathcal O$ equals $\mathcal M_x$ and $\mathcal M_y$. The symmetries $\{ \mathcal C_{2z} | 000 \}$ and $\{ \mathcal O | \tfrac12 \tfrac12 \tfrac12 \}$ satisfy
\be
\{ \mathcal O | \tfrac12 \tfrac12 \tfrac12 \}  \{ \mathcal C_{2z} | 000 \} = \{ \mathcal O \mathcal C_{2z} | \tfrac12 \tfrac12 \tfrac12 \} =\mathcal R  \{ \mathcal C_{2z} \mathcal O | \tfrac12 \tfrac12 \tfrac12 \} = \mathcal R  \{ \mathbb{1} | 110 \}  \{ \mathcal C_{2z} | 000 \}  \{ \mathcal O | \tfrac12 \tfrac12 \tfrac12 \},
\ee
which implies that their representations anticommute on the lines defined by $k_x=k_y=0$ and $k_x=k_y=\pi$. This anticommutation relation implies a twofold degeneracy of all bands on the vertical rotation lines. Now consider the product $\mathcal T\{ \mathcal I | 000 \}$, which also leaves these lines invariant. To see whether $\mathcal T\{ \mathcal I | 000 \}$ promotes the twofold degeneracy to fourfold degeneracy, as it did in the case of the $(k_x,k_z) = (\pi,0)$ line, we must examine the square of $\mathcal T\{ \mathcal I | 000 \} \{ \mathcal O | \tfrac12 \tfrac12 \tfrac12 \}$. For $\mathcal O=M_x$ we can refer to \eqref{T.I.M_x_squared}; for $\mathcal O=M_y$ we have
\be
(\mathcal T\{ \mathcal I | 000 \}\{ \mathcal M_y | \tfrac12 \tfrac12 \tfrac12 \} )^2 = (\mathcal T\{ \mathcal C_{2y} | \tfrac{\bar 1}{2} \tfrac{\bar 1}{2}  \tfrac{\bar 1}{2}  \} )^2 =\mathcal T^2 \mathcal R \{ \mathbb{1} | 0\bar 10 \}.
\ee
In both cases we find that $\mathcal T\{ \mathcal I | 000 \} \{ \mathcal O | \tfrac12 \tfrac12 \tfrac12 \}$ squares to $+1$ on the $k_x=k_y=0$ lines, and to $-1$ on the $k_x=k_y=\pi$ line. As a result, all bands on the $k_x=k_y=\pi$ are manifestly fourfold degenerate in the presence of these symmetries.

\end{itemize}

\subsection{Analysis of Hamiltonian for N\'eel vector $\bN \parallel \hat z$ }

We now analyze the generic features of the 3D rutile model by focusing specifically on planes, lines, and points of high symmetry. All analysis will assume that the N\'eel vector points in the $z$ direction, i.e., $\bN = N\hat z$, such that $\mathcal H_{\text{AM}} (\bk) $ is given by \eqref{H_AM_3D_SOC} (with $J'=0$). We examine three features in particular: (i) mirror symmetry protected Weyl line nodes on the $k_z=\pi$ plane, (ii) the fate of the symmetry-enforced fourfold Dirac points at the high symmetry point $Z$ in the presence of altermagnetism, and (iii) the fate of the Dirac line nodes on the $MX$ and $MY$ lines in the presence of altermagnetism.

\begin{itemize}

\item {\it Weyl line nodes on the $k_z=\pi$ plane.}

In Sec.~\ref{ssec:ham-analysis-3D}  we showed that, in the absence of SOC, Weyl line nodes exist in the $k_z=\pi$ plane, and that these line nodes are protected by mirror symmetry. Here we examine the stability of these line nodes in the presence of SOC. The first observation is that a N\'eel vector along the $z$ direction, i.e., $\bN \parallel \hat z$, preserves the mirror symmetry $\{ \mathcal M_{z} |000 \} $. This implies that line nodes can be locally stable if they originate from a crossing of bands with opposite mirror eigenvalues. In Sec.~\ref{ssec:ham-analysis-3D}, we determined that, without SOC, the line nodes come from the crossing of bands with the same spin but opposite sublattice index. To get a sense for whether such crossings remain protected, consider the representation of the mirror operator on the Bloch states in the $k_z=\pi$ plane, which in the presence of SOC is given by
\be
U_{\mathcal M_z} = -i\tau_z\sigma_z. \label{mirror_Mz_3D}
\ee
This form indicates that states with the same spin but opposite sublattice index have opposite mirror eigenvalue. We therefore expect the line nodes to remain stable as SOC is included in the Hamiltonian. 

To examine this in detail, consider the Hamiltonian on the $k_z=\pi$ plane, which is given by
\be
\mathcal H (k_x,k_y,\pi)  =   2 t_2( 1- c_x-c_y  ) \tau_0 -  4 t_d s_xs_y \tau_z + \lambda_2 (s_{x/2}c_{y/2}\tau_y \sigma_x - s_{y/2}c_{x/2}\tau_y \sigma_y)+ JN \tau_z \sigma_z  \label{H_kz_pi_plane}
\ee
Note that the SOC term proportional to $\lambda_1$ vanishes on the $k_z=\pi$ plane. Since the Hamiltonian must commute with $U_{\mathcal M_z}$, it can be block diagonalized by expressing it in basis of eigenstates of $U_{\mathcal M_z}$. This results in two $2\times 2$ blocks labeled by the mirror eigenvalues $\pm i$. To express the $2\times 2$ Hamiltonian within each mirror sector, we introduce a set of Pauli matrices $(\mu_x,\mu_y,\mu_z)$ and find that the blocks $\mathcal H^\pm (\bk) $ are given by
\be
\mathcal H^\pm (\bk)   = [2 t_2( 1- c_x-c_y  ) \mp J N ] \tau_0 - 4 t_d s_xs_y \mu_z +  \lambda_2   (s_{x/2} c_{y/2}  \mu_y \mp  s_{y/2} c_{x/2} \mu_x ) ,\label{H_mirror_kz_pi_plane}
\ee
where $\pm$ corresponds to the mirror eigenvalues $\pm i$. It is straightforward to obtain the energies in each mirror sector and we find
\be
\mathcal E^{m}_{\bk, \pm} =2 t_2( 1- c_x-c_y  )- m JN  \pm \sqrt{( 4 t_d)^2 s^2_xs^2_y + \lambda^2_2 (1-c_{x} c_{y})/2 } \label{E_kz_pi_plane},
\ee
where $m=\pm$ denotes the mirror sectors. 

To proceed, we make the assumption that $JN>0$ (the case $JN<0$ is analogous). To demonstrate the presence of line nodes, we consider possible crossings of bands from different mirror sectors. Given the assumption $JN>0$, such crossings must occur between the bands $\mathcal E^{+}_{\bk, +}$ and $\mathcal E^{-}_{\bk, -}$. The condition for a crossing of these two bands occur reads as
\be
JN  = \sqrt{( 4 t_d)^2 s^2_xs^2_y + \lambda^2_2 (1-c_{x} c_{y})/2 }  \label{line-node-condition}
\ee
which defines one constraint on a set of two variables $(k_x,k_y)$, and is therefore generically satisfied on a line (i.e. a 1D manifold) on the $k_z=\pi$ plane. In particular, we have seen in Sec.~\ref{ssec:ham-analysis-3D} that such crossings indeed occur when $\lambda_2=0$, as long as $JN < 4|t_d|$. It is clear from Eq.~\eqref{line-node-condition} that including SOC by activating $\lambda_2$ does not remove the line nodes, but simply changes their contour on the $k_z=\pi$ plane (see the main text). It is also clear from Eq.~\eqref{line-node-condition} that increasing $JN$ beyond a critical value (which is different from $JN < 4|t_d|$ in the presence of SOC) will remove the line nodes, as \eqref{line-node-condition} no longer has solutions. We conclude this discussion by stressing that in the presence of SOC the line nodes can no longer be associated with spin, as was possible in the absence of SOC.

\item {\it Weyl line nodes from an expansion around the $Z$ point (3D Dirac point).}

To gain further insight into the presence of line nodes on the $k_z=\pi$ plane, it is instructive to expand the Hamiltonian $\mathcal H_0 + \mathcal H_{\text{SO}}$ around the $Z$ point with momentum $\bQ_Z = (0,0,\pi)$, and then examine the effect of $\mathcal H_{\text{AM}}$. Hence, in this approach, the full Hamiltonian including SOC (but without altermagnetism) is expanded around a point of high symmetry, and then the effect of magnetism is considered. The key observation that motives this approach is the symmentry-enforced presence of a fourfold degenerate Dirac point at $Z$.

To expand the Hamiltonian around the $Z$ point, we first expand the lattice harmonics; this yields 
\be
s_{\bQ_Z+\bq } = -4q_z , \qquad  s'_{\bQ_Z+\bq } = -  2 , \qquad d_{\bQ_Z+\bq } = 4 q_x q_y, 
\ee
for the lattice harmonics already present in the Hamiltonian without SOC. To expand the SOC terms we define the harmonics
\be
 \bar d_{\bk} = c_{x/2}c_{y/2}c_{z/2} (c_x-c_y), \qquad (\delta^1_\bk, \delta^2_\bk) = ( s_{z/2} s_{x/2}c_{y/2}, s_{z/2} s_{y/2}c_{x/2}),
\ee
and expand around the $Z$ point to obtain
\be
 \bar d_{\bQ_Z+\bq} = \frac14 q_z (q^2_x-q^2_y), \qquad (\delta^1_{\bQ_Z+\bq}, \delta^2_{\bQ_Z+\bq }) = \frac12 (q_x, q_y). 
\ee
Substituting these expansions back into the Hamiltonian and retaining only terms linear in $\bq$, we obtain
\be
\mathcal H(\bQ_Z+\bq) =v q_x \tau_y\sigma_x -v  q_y \tau_y\sigma_y+ v_z  q_z \tau_x  , \label{eq:aux_HZ}
\ee
which is the Hamiltonian of a 3D Dirac point with anisotropic velocities
\be
v = \frac12 \lambda_2, \qquad v_z = 4t_1.
\ee
It is important to stress that the $Z$ point is a Dirac point only when SOC is included. To proceed, we now consider the effect of the altermagnetic term. The Dirac Hamiltonian in the presence of altermagnetism reads as
\be
\mathcal H(\bQ_Z+\bq) =v q_x \tau_y\sigma_x -v  q_y \tau_y\sigma_y+ v_z  q_z \tau_x  + JN \tau_z \sigma_z
\ee
which is straightforwardly diagonalized to obtain the energies. The energies follow from the equation
\be
\mathcal E^2 = \left(\sqrt{v^2 q^2_x + v^2 q^2_y} \pm JN\right)^2 + v_z^2 q_z^2,
\ee
which indeed defines four spectral branches. The spectrum is gapped when $q_z\neq 0$, but has line nodes when $q_z=0$. This follows from the fact that the equation $(\sqrt{v^2 q^2_x + v^2 q^2_y} \pm JN )^2=0 $ is guaranteed to have solutions, and more specifically is guaranteed to have solutions on a circle in $(q_x,q_y)$ space. The radius of the circle is given by $q_* = |JN / v|$. 

The presence of line nodes may be understood in slightly different way by performing a basis transformation of the Hamiltonian using the matrix
\be
V = \begin{pmatrix} 1 & 0 &0 &0 \\ 0 &0& 1& 0\\0 &0 & 0 &  1 \\0 &1 & 0 &  0 \end{pmatrix}.
\ee
Restricting to momenta $\widetilde{\bq} = (q_x, q_y, 0)$, we find
\be
V^\dagger \mathcal H(\bQ_Z+\widetilde{\bq}) V = \begin{pmatrix}JN & v( q_y -i q_x )&0 &0 \\ v( q_y +i q_x ) & JN& 0& 0\\0 &0 & -JN &  -v( q_y +i q_x ) \\0 &0 &  -v( q_y -i q_x )&  -JN \end{pmatrix}
\ee
which shows that the Hamiltonian on the $q_z=0$ plane can be interpreted as two Dirac points shifted in energy. This will necessarily lead to crossings of energy bands on a circle surrounding $(q_x,q_y) = (0,0)$.

Note that Eq. \ref{eq:aux_HZ}  is independent of the hopping anisotropy $t_d$ because we restricted our analysis to terms that are linear in momentum only. Reinstating the effect of anisotropic hopping proportional to $t_d$ gives a continuum Hamiltonian of the form
\be
\mathcal H(\bQ_Z+\bq) =v_x q_x \tau_y\sigma_x +v_y  q_y \tau_y\sigma_y+ v_z  q_z \tau_x -4   t_d q_xq_y \tau_z + JN \tau_z \sigma_z,
\ee
which contains an additional term quadratic in the momentum. Setting $q_z=0$ then yields energies obtained from the equation
\be
\mathcal E^2 = \left[\sqrt{v^2 ( q^2_x + q^2_y)+ (4t_d)^2 q^2_xq^2_y} \pm JN\right]^2 ,
\ee
which shows that the line nodes remain present but are no longer perfectly circular. 

A similar analysis can be performed around the $A$ point, given by $\bQ_A = (\pi, \pi,\pi)$. In the nonmagnetic but spin-orbit coupled system the $A$ point is not a Dirac point, however. Nonetheless, a similar analysis yields Weyl line nodes surrounding the $A$ point when magnetism is included.

\item {\it Splitting Dirac line nodes into Weyl line nodes on the $MX$ and $MY$ lines.}

In this final part of our analysis we focus on the $MX$ and $MY$ lines. The goal is to show that these lines realize Weyl nodal lines, which may be viewed as Dirac nodal lines split by altermagnetism. Since the $MX$ and $MY$ lines are related by symmetry, it is sufficient to consider one of these; we choose $MX$. 

Similar to the previous analysis, we expand around $(k_x,k_z) = (\pi,0)$ in small momenta $(q_x,q_z)$. The lattice harmonics are expanded as
\be
s_{\pi+q_x,k_y,q_z} = -4q_x c_{y/2} , \qquad  s'_{\pi+q_x,k_y,q_z} =   2c_y , \qquad d_{\pi+q_x,k_y,q_z} = -4 q_x s_y, 
\ee
and
\be
 \bar d_{\pi+q_x,k_y,q_z} =  \frac12 q_x (1+c_y), \qquad \delta^1_{\pi+q_x,k_y,q_z}= \frac12 q_z c_{y/2}. 
\ee
Note that since here we expand on a line we keep the full dependence on $k_y$, which parametrizes the line. The expanded Hamiltonian in the absence of altermagnetism takes the form
\be
\mathcal H(\pi+q_x,k_y,q_z) = -2 t_2c_y \tau_0 +  q_x \left[ 4t_1 c_{y/2} \tau_x + 4t_d s_y \tau_z +\frac12 \lambda_1(1+c_y)\tau_y\sigma_z \right] + \frac12\lambda_2 q_z c_{y/2} \tau_y\sigma_x,
\ee
which is indeed recognized as describing a Dirac nodal line as a function of $k_y$. When $(q_x,q_z)=(0,0)$ all energies are fourfold degenerate and the splitting away from the $(k_x,k_z) = (\pi,0)$ line is linear in $(q_x,q_z)$. Note that the velocities are a function of $k_y$. 

In the presence of altermagnetism the fourfold degenerate energy level on the $(k_x,k_z) = (\pi,0)$ line is split, yielding two sets of two states pushed up/down in energy by an amount $\pm JN$. Projecting the expanded Hamiltonian into these two subspaces and ignoring the coupling between the subspaces produces a Hamiltonian given by
\be
\mathcal H^{\pm JN}(\pi+q_x,k_y,q_z) =  \begin{pmatrix}\pm JN + \tilde v_x q_x & -i \tilde v_z q_z\\ i \tilde v_z q_z &  \pm JN -  \tilde v_x q_x\end{pmatrix}, \qquad  \tilde v_x = 4t_d s_y, \qquad   \tilde v_z = \frac12\lambda_2 c_{y/2}. 
\ee
Each Hamiltonian describes a Weyl line node with linear dispersion away from $(k_x,k_z) = (\pi,0)$ and $k_y$-dependent velocities.

\end{itemize}

\bibliography{main-mirror-chern-bands-altermagnets}